\documentclass[prc,aps,eqsecnum,amssymb,floatfix,showpacs]{revtex4}
\usepackage{dcolumn}
\usepackage{bm}
\usepackage{graphicx}
\usepackage{morefloats}

\begin{document}
\title{Inclusive neutrino scattering off deuteron from threshold to GeV energies}
\author{G.\ Shen$^{\,{\rm a}}$, L.E.\ Marcucci$^{\,{\rm b,c}}$, J.\ Carlson$^{\,{\rm a}}$,
S.\ Gandolfi$^{\,{\rm a}}$, and R.\ Schiavilla$^{\,{\rm d,e}}$}

\affiliation{
$^{\rm a}$\mbox{Theoretical Division, Los Alamos National Laboratory, NM 87545, USA}\\
$^{\,{\rm b}}$\mbox{Department of Physics, University of Pisa, 56127 Pisa, Italy}\\
$^{\,{\rm c}}$\mbox{INFN-Pisa, 56127 Pisa, Italy}\\
$^{\rm d}$\mbox{Department of Physics, Old Dominion University, Norfolk, VA 23529, USA}\\
$^{\rm e}$\mbox{Jefferson Lab, Newport News, VA 23606, USA}\\
}
\date{\today}

\begin{abstract}
Background: Neutrino-nucleus quasi-elastic scattering is crucial to interpret the neutrino oscillation results in long baseline neutrino experiments. There are rather large uncertainties in the cross section, due to insufficient knowledge on the role of two-body weak currents. 
Purpose: Determine the role of two-body weak currents in neutrino-deuteron quasi-elastic scattering up to GeV energies. 
Methods: Calculate cross sections for inclusive neutrino scattering off deuteron induced by neutral and charge-changing weak currents, from threshold up to GeV energies, using the Argonne $v_{18}$ potential and consistent nuclear electroweak currents with one- and two-body terms. 
Results: Two-body contributions are found to be small, and increase the cross sections obtained with one-body currents by less than 10\% over the whole range of energies. Total cross sections obtained by describing the final two-nucleon states with plane waves differ negligibly, for neutrino energies $\gtrsim 500$ MeV, from those in which interaction effects in these states are fully accounted for. The sensitivity of the calculated cross sections to different models for the two-nucleon potential and/or two-body terms in the weak current is found to be weak. Comparing cross sections to those obtained in a naive model in which the deuteron is taken to consist of a free proton and neutron at rest, nuclear structure effects are illustrated to be non-negligible. 
Conclusion: Contributions of two-body currents in neutrino-deuteron quasi-elastic scattering up to GeV are found to be smaller than 10\%. Finally, it should be stressed that the results reported in this work do not include pion production channels. 

\end{abstract}

\pacs{25.10.+s, 25.30.Pt}

\maketitle
\section{Introduction}
\label{sec:intro}
In last few years, inclusive neutrino scattering from nuclear targets
has become a hot topic.  Interest has been spurred by the anomaly
observed in recent neutrino quasi-elastic scattering data on
$^{12}$C~\cite{Aguilar08,Butkevich10}, i.e.~the excess, at relatively low
energy, of measured cross section relative to theoretical calculations.
Analyses based on these calculations have led to speculations that our
present understanding of the nuclear response to charge-changing weak
probes may be incomplete~\cite{Benhar10}, and, in particular, that the
momentum transfer dependence of the axial form factor of the nucleon,
specifically the cutoff value of its dipole parameterization~\cite{Juszczak10},
may be quite different from that obtained from analyses of pion electro-production
data~\cite{Amaldi79} and measurements of the reaction $n(\nu_\mu,\mu^-)p$ in the
deuteron at quasi-elastic kinematics~\cite{Baker81,Kitagaki83} and
of $\nu_\mu p$ and $\overline{\nu}_\mu p$ elastic scattering~\cite{Ahrens87}
($\Lambda_A \simeq 1.20$ GeV versus $\Lambda_A \simeq 1$ GeV).
However, it should be emphasized that the calculations on which these
analyses are based use rather crude models of nuclear structure---Fermi
gas or local density approximations of the nuclear matter spectral function---as
well as simplistic treatments of the reaction mechanism, and
should therefore be viewed with skepticism.

In this paper, we calculate cross sections for inclusive neutrino scattering off
deuteron in a wide energy range, from threshold up to 1 GeV.  The motivations
for undertaking such a work are twofold. The first is to provide a benchmark
for studies of electro-weak inclusive response in light nuclei  we intend to carry
out in the near future.  The second motivation has to do with plans~\cite{Garvey12},
still under development, to determine the neutrino flux in accelerator-based experiments from measurements of inclusive
cross sections on the deuteron.  In particular, in charged-current neutrino capture on deuteron,
the final states $ppl^{-}$ can be measured, in principle, very well.  Clearly, accurate predictions
for these cross sections are crucial for a reliable determination of the flux. 

A number of studies of neutrino-deuteron scattering at low and intermediate energies ($\lesssim 150$ MeV)
were carried out in the past decades, see Ref.~\cite{Kubodera94} for a review of work done up to
the mid 1990's.  These efforts culminated in the Nakamura {\it et al.}'s 2001 and 2002 calculations
of the cross sections for neutrino disintegration of the deuteron induced by neutral and charge-changing
weak currents.  These calculations were based on bound- and scattering-state wave functions
obtained from last-generation realistic potentials, and used a realistic model for the nuclear
weak current, including one- and two-body terms.  The vector part of this current was shown to provide
an excellent description of the $np$ radiative capture cross section for neutron energies up to 100
MeV~\cite{Nakamura01}, while the axial part was constrained to reproduce the Gamow-Teller matrix
element in tritium $\beta$-decay~\cite{Nakamura02}.  The Nakamura {\it et al.} studies have played
an important role in the analysis and interpretation of the Sudbury Neutrino Observatory (SNO)
experiments~\cite{Ahmad02}, which have established solar neutrino oscillations and the validity
of the standard model for the generation of energy and neutrinos in the sun~\cite{Bahcall04}.

In the present work, we use the same theoretical framework as the authors of
Refs.~\cite{Nakamura01,Nakamura02}, but include refinements in the modeling of the weak
current---which however, as shown in Sec.~\ref{sec:res}, will turn out to have a minor impact on the
predicted cross sections---and extend the range of neutrino energies up to 1 GeV.  While the
theoretical approach is essentially the same, nevertheless the way in which the calculations are
carried out in practice is rather different from that used in those earlier papers, which relied
on a multipole expansion of the weak transition operators, and evaluated the cross section by
summing over a relatively large number of final two-nucleon channels states.  In contrast, we evaluate,
by direct numerical integrations, the matrix elements of the weak current between the deuteron
and the two-nucleon scattering states labeled by the relative momentum ${\bf p}$ (and in given
pair spin and isospin channels), thus avoiding cumbersome multipole expansions.   Differential
cross sections are then obtained by integrating over ${\bf p}$ (and summing over the discrete
quantum numbers) appropriate combinations of these matrix elements, i.e.~by calculating
the weak response functions.  The techniques developed here for the deuteron should prove
valuable when we will attempt the Green's function Monte Carlo calculation of these response
functions (or rather, their Laplace transforms~\cite{Carlson92}) in $A>2$ nuclei.

This paper is organized as follows.  In Sec.~\ref{sec:xsect} and Appendix~\ref{app:app1} we
present the neutrino and antineutrino differential cross sections expressed in terms of response functions,
while in Sec.~\ref{sec:cnts} we provide a succinct description of the neutral and charge-changing
weak-current model.  In Sec.~\ref{sec:calc} we outline the methods used to obtain the two-nucleon
bound and continuum states, and discuss the numerical evaluation of the response functions.
A variety of results for the neutral current processes $^2$H($\nu_l,\nu_l$)$pn$ and
$^2$H($\overline{\nu}_l,\overline{\nu}_l$)$pn$, and charge-changing processes $^2$H($\nu_e,e^-$)$pp$ and
$^2$H($\overline{\nu}_e,e^+$)$nn$ are presented in Sec.~\ref{sec:res}, including the
sensitivity of the calculated cross sections to (i) interaction effects in the final states, 
(ii) different short-range behaviors of the two-body axial weak currents, and (iii) different
potential models to describe the two-nucleon bound and continuum states.  In order
to illustrate the effects of nuclear structure, we compare these cross sections to those
obtained in a naive model in which the deuteron is taken to consist of a free proton
and neutron (the free nucleon cross sections are listed for reference
in Appendix~\ref{app:app2}).  Concluding remarks and an outlook are given in Sec.~\ref{sec:concl}.
\section{Inclusive neutrino scattering off deuteron}
\label{sec:xsect}
The differential cross section for neutrino ($\nu$) and
antineutrino ($\overline{\nu}$) inclusive scattering off deuteron, specifically
the processes
\begin{equation}
\nu_l +d \longrightarrow \nu_l+p+n \ , \qquad  \overline{\nu}_l+d \longrightarrow  \overline{\nu}_l+p+n 
\label{eq:nnr}
\end{equation}
induced by neutral weak currents (NC), and the processes
\begin{equation}
\nu_l +d \longrightarrow l^- +p+p \  , \qquad  \overline{\nu}_l+d \longrightarrow l^+ +n+n 
\label{eq:ncr}
\end{equation} 
induced by charge-changing weak currents (CC), can be expressed as
\begin{eqnarray}
\left(\frac{ {\rm d}\sigma}{ {\rm d}\epsilon^\prime {\rm d}\Omega}\right)_{\nu/\overline{\nu}}&=& \frac{G^2}{2\pi^2}\, 
k^\prime \epsilon^\prime \,F(Z,k^\prime)\, {\rm cos}^2 \frac{\theta}{2}\Bigg[
R_{00} +\frac{\omega^2}{ q^2}\, R_{zz} - \frac{\omega}{q} R_{0z} +
\left( {\rm tan}^2\frac{\theta}{2}+\frac{Q^2}{2\, q^2}\right) R_{xx+yy} \nonumber \\
&&\mp {\rm tan}\frac{\theta}{2}\, \sqrt{  {\rm tan}^2\frac{\theta}{2}+\frac{Q^2}{ q^2} } \, R_{xy}
\Bigg] \ ,
\label{eq:xsw}
\end{eqnarray}
where $G$=$G_F$ for the NC processes and $G$=$G_F \, {\rm cos}\, \theta_C$ for the
CC processes, and the $-$ ($+$) sign in the last term is relative to the $\nu$ ($\overline{\nu}$)
initiated reactions.  Following Ref.~\cite{Nakamura02}, we adopt the value
$G_F=1.1803\times 10^{-5}$ GeV$^{-2}$ as obtained from an analysis of
super-allowed $0^+ \rightarrow 0^+$ $\beta$-decays~\cite{Towner99}---this
value includes radiative corrections---while ${\rm cos}\, \theta_C$ is taken as 0.97425
from~\cite{PDG}.  The initial neutrino four-momentum is $k^\mu=(\epsilon, {\bf k})$, the final
lepton four momentum is $k^{\mu \,\prime}=(\epsilon^\prime,{\bf k}^\prime)$, and the
lepton scattering angle is denoted by $\theta$.  We have also defined
the lepton energy and momentum transfers as $\omega=\epsilon-\epsilon^\prime$
and ${\bf q}={\bf k}-{\bf k}^\prime$, respectively, and the squared four-momentum
transfer as $Q^2=q^2-\omega^2 > 0$.  The Fermi function $F(Z,k^\prime)$ with $Z=2$
accounts for the Coulomb distortion of the final lepton wave function in the the charge-raising
reaction,
\begin{equation}
F(Z,k^\prime) = 2\, (1+\gamma)\, (2\, k^\prime\, r_d)^{2\,\gamma-2}\, {\rm exp} \left(\pi\, y\right)\,
\Bigg| \frac{\Gamma(\gamma+i\, y)}{\Gamma(1+2\,\gamma)} \Bigg|^2 \ , \qquad
\gamma=\sqrt{1-\left(Z\,\alpha\right)^2} \ ,
\end{equation}
it is set to one otherwise.  Here $y = Z\, \alpha \, \epsilon^\prime/k^\prime$,
$\Gamma(z)$ is the gamma function, $r_d$ is the deuteron radius
($r_d=1.967$ fm), and $\alpha$ is the fine structure constant.
Radiative corrections for the CC and NC processes due to bremsstrahlung and
virtual photon- and $Z$-exchanges have been evaluated by the
authors of Refs.~\cite{Towner98,Kurylov02} at the low energies
($\sim 10$ MeV) relevant for the SNO experiment, which measured the neutrino
flux from the $^8$B decay in the sun.  These corrections are neglected
in the present work, since its focus is on scattering of neutrinos with energies
larger than 100 MeV.  We are not (or not yet, at least) concerned with
providing cross section calculations with \% accuracy in this regime.
Lastly, the nuclear response functions are defined as
\begin{eqnarray}
\label{eq:r1}
R_{00}(q,\omega) &=&\frac{1}{3} \sum_{M } \sum_f \delta( \omega+m_d-E_f)\,
\langle f\mid j^0({\bf q},\omega) \mid d, M \rangle\langle f\mid j^0({\bf q},\omega) \mid d, M \rangle^*  \ ,\\
\label{eq:r2}
R_{zz}(q,\omega) &=&\frac{1}{3} \sum_{M } \sum_f \delta( \omega+m_d-E_f)\,
\langle f\mid j^z({\bf q},\omega) \mid d, M \rangle\langle f\mid j^z({\bf q},\omega) \mid d, M \rangle^*\ ,  \\
\label{eq:r3}
R_{0z}(q,\omega) &=&\frac{2}{3} \sum_{M } \sum_f \delta( \omega+m_d-E_f)\, {\rm Re}\Big[
\langle f\mid j^0({\bf q},\omega) \mid d, M \rangle\langle f\mid j^z({\bf q},\omega) \mid d, M \rangle^*\Big]  \ ,\\
\label{eq:r4}
R_{xx+yy}(q,\omega) &=&\frac{1}{3} \sum_{M } \sum_f \delta( \omega+m_d-E_f)\,\Big[
\langle f\mid j^x({\bf q},\omega) \mid d, M \rangle\langle f\mid j^x({\bf q},\omega) \mid d, M \rangle^* \nonumber \\
&&\qquad \qquad+\langle f\mid j^y({\bf q},\omega) \mid d, M \rangle\langle f\mid j^y({\bf q},\omega) \mid d, M \rangle^*  \Big] \ ,\\
R_{xy}(q,\omega) &=&\frac{2}{3} \sum_{M } \sum_f \delta( \omega+m_d-E_f)\, {\rm Im}\Big[
\langle f\mid j^x({\bf q},\omega) \mid d, M \rangle\langle f\mid j^y({\bf q},\omega) \mid d, M \rangle^* \Big] \ ,
\label{eq:r5}
\end{eqnarray}
where $\mid \!d, M\rangle$ and $\mid \!\! f\rangle$ represent, respectively, the initial deuteron state in spin projection
$M$ and the final two-nucleon state of energy $E_f$, and $m_d$ is the deuteron
rest mass.  The three-momentum
transfer ${\bf q}$ is taken along the $z$-axis (i.e., the spin-quantization axis), and
 $j^\mu({\bf q},\omega)$ is the time component (for $\mu=0$) or space
component (for $\mu=x,y,z$) of the NC or CC.

The expression above for the CC cross section is valid in the limit $\epsilon^\prime \simeq
k^\prime$, in which the lepton rest mass is neglected.   At small incident neutrino energy,
this approximation is not correct.  Inclusion of the lepton rest mass leads to
changes in the kinematical factors multiplying the various response functions.  The resulting
cross section is given in Appendix~\ref{app:app1}.

\section{Neutral and charge-changing weak currents}
\label{sec:cnts}
We denote the neutral and charge-changing weak currents
as $j^\mu_{NC}$ and $j^\mu_{CC}$, respectively.  The former is
given by
\begin{equation}
j^\mu_{NC}=-2\, {\rm sin}^2\theta_W\, j^\mu_{\gamma, S} + (1-2\, {\rm sin}^2\theta_W) \, j^\mu_{\gamma, z} 
+\, j^{\mu5}_z \ ,
\end{equation}
where $\theta_W$ is the Weinberg angle (${\rm sin}^2\theta_W=0.2312$~\cite{PDG}), $j^\mu_{\gamma,S}$
and $j^\mu_{\gamma,z}$ denote, respectively,
the isoscalar and isovector pieces of the electromagnetic current, and $j^{\mu5}_z$ denotes the isovector
piece of the axial current (the $z$ on the isovector terms indicates that they transform as the $z$-component
of an isovector under rotations in isospin space).  Isoscalar contributions to $j^\mu_{NC}$
associated with strange quarks are ignored, since experiments at Bates~\cite{Spayde00} and JLab~\cite{Ahmed12}
have found them to be very small.

The charge-changing weak current is written as the sum of polar- and axial-vector components
\begin{equation}
j^\mu_{CC}=j^\mu_{\pm}+j^{\mu5}_{\pm} \ , \qquad j_\pm = j_x \pm i\, j_y \ .
\end{equation}
The conserved-vector-current (CVC) constraint relates the polar-vector components $j^\mu_b$ of the
charge-changing weak current to the
isovector component $j^\mu_{\gamma,z}$ of the electromagnetic current via
\begin{equation}
\left[ \, T_a \, , \, j^\mu_{\gamma,z} \, \right]=i\, \epsilon_{azb}\, j^\mu_b \ ,
\end{equation}
where $T_a$ are isospin operators.  We now turn to a discussion of the one- and two-body
contributions to the NC and CC.
\subsection{One-body terms}
\label{sec:1b}
The isoscalar components of the one-body electromagnetic current
are given by
\begin{eqnarray}
j^0_{\gamma,S}(i)&=& \left[ \frac{G_E^S(Q^2)}{2\, \sqrt{1+Q^2/(4\, m^2)} } 
-i\, \frac{2\, G_M^S(Q^2)-G_E^S(Q^2)}{8\, m^2}
{\bf q} \cdot \left( {\bm \sigma}_i \times {\bf p}_i\right) \right] \, {\rm e}^{i\, {\bf q}\cdot {\bf r}_i}\ , \\
{\bf j}^\perp_{\gamma, S}(i)&=&\left[  \frac{G_E^S(Q^2)}{2\, m} \, {\bf p}^\perp_i  
-i\, \frac{G_M^S(Q^2)}{4\, m}\, {\bf q}\times  {\bm \sigma}_i  \right] \, {\rm e}^{i\, {\bf q}\cdot {\bf r}_i} \ , \\
j^\parallel_{\gamma, S}(i)&=& \frac{\omega}{q} \,  j^0_{\gamma,S}(i) \ ,
\label{eq:jlong}
\end{eqnarray}
and the corresponding isovector components of $j^\mu_{\gamma, z}$ are obtained by the replacements
\begin{equation}
G_E^S(Q^2) \longrightarrow G_E^V(Q^2)\, \tau_{i,z} \ , \qquad G_M^S(Q^2) \longrightarrow
G_M^V(Q^2)\, \tau_{i,z} \ ,
\end{equation}
where $G_E^{S/V}$ and $G_M^{S/V}$ are the isoscalar/isovector combinations of the proton
and neutron electric ($E$) and magnetic ($M$) form factors, ${\bf r}_i$ and ${\bf p}_i$ are
the position and momentum operators of nucleon $i$,
${\bm \sigma}_i$ and $\tau_{i,z}$ are its Pauli spin and isospin matrices, and $m$ is the nucleon
mass (0.9389 GeV).  Note that
we have decomposed ${\bf j}_{\gamma, S}$ and ${\bf j}_{\gamma,z}$ into
transverse ($\perp$) and longitudinal ($\parallel$) components to the momentum
transfer ${\bf q}$, and have used
current conservation to relate the latter to the
isoscalar and isovector charge operators $j^0_{\gamma, S}$ and $j^0_{\gamma,z}$.
The isovector components of the axial weak neutral current $j^{\mu5}_z$ are
given by
\begin{eqnarray}
\label{eq:rho5}
j^{05}_z(i) &=& -\frac{G_A(Q^2)}{4\, m}\, \tau_{i,z}\, {\bm \sigma}_i\,  \cdot \left[\, {\bf p}_i\, , \, 
{\rm e}^{i\, {\bf q}\cdot {\bf r}_i} \right]_+ \ ,\\
\label{eq:j5}
{\bf j}^{5}_z(i)&=& -\frac{G_A(Q^2)}{2}\, \tau_{i,z}\, \Bigg [ {\bm \sigma}_i\,  {\rm e}^{i\, {\bf q}\cdot {\bf r}_i} 
-\frac{1}{4\, m^2} \Bigg( {\bm \sigma}_i \left[ \,{\bf p}_i^2 \, , \, {\rm e}^{i\, {\bf q}\cdot {\bf r}_i} \right]_+
- \left[ ({\bm \sigma}_i \cdot {\bf p}_i) \, {\bf p}_i \, , \, {\rm e}^{i\, {\bf q}\cdot {\bf r}_i} \right]_+ \nonumber \\
&&-\frac{1}{2} {\bm \sigma}_i \cdot {\bf q} \left[ \, {\bf p}_i \, , \, {\rm e}^{i\, {\bf q}\cdot {\bf r}_i} \right]_+
-\frac{1}{2} {\bf q} \left[ \,{\bm \sigma}_i \cdot  {\bf p}_i \, , \, {\rm e}^{i\, {\bf q}\cdot {\bf r}_i} \right]_+
+i\, {\bf q}\times {\bf p}_i\,  {\rm e}^{i\, {\bf q}\cdot {\bf r}_i}\Bigg)  \Bigg] \ ,
\end{eqnarray}
where $G_A$ is the nucleon axial form factor, and $\left[ \dots\, , \, \dots\right]_+$ denotes the anticommutator.
The operators above include terms of order $(v/c)^2$ in the non-relativistic expansion of
the single-nucleon covariant currents.  These have been neglected in the study
of Ref.~\cite{Nakamura02}.  The proton and neutron electromagnetic and nucleon axial form factors
are  parametrized as
\begin{eqnarray}
G_E^p(Q^2) &=& G_D(Q^2)  \ ,  \qquad G_E^n(Q^2) =-\mu_n\, \frac{Q^2}{4\, m^2} \frac{G_D(Q^2)}{1+Q^2/m^2} \ ,\\
G_M^p(Q^2)&= &\mu_p \, G_D(Q^2)  \ ,  \qquad G_M^n(Q^2)= \mu_n\, G_D(Q^2) \ ,  \\
G_D(Q^2)&=&\frac{1}{\left(1+Q^2/\Lambda^2\right)^2} \ ,  \qquad G_A(Q^2) = \frac{g_A}{\left(1+Q^2/\Lambda_A^2\right)^2} \ ,
\label{eq:gga}
\end{eqnarray}
from which the isoscalar and isovector combinations are obtained as $G_{E,M}^{S/V}=G^p_{E,M} \pm G^n_{E,M}$.
The proton and neutron magnetic moments are $\mu_p=2.793$ and $\mu_n=-1.913$ in units of
nuclear magnetons (n.m.), and the nucleon axial-vector coupling constant
is taken to be $g_A=1.2694$~\cite{PDG}.  The values for the cutoff masses $\Lambda$ and $\Lambda_A$
used in this work are 0.833 GeV and 1 GeV, respectively.   The former is from fits to
elastic electron scattering data off the proton and deuteron~\cite{Hyde04}, while the latter is 
from an analysis of pion electroproduction~\cite{Amaldi79} and neutrino
scattering~\cite{Baker81,Kitagaki83,Ahrens87} data.
Uncertainties in the $Q^2$ dependence of the axial form factor,
in particular the value of $\Lambda_A$, could significantly impact predictions
for the neutrino cross sections under consideration.  As mentioned earlier,
recent analyses of neutrino quasi-elastic scattering data on nuclear targets~\cite{Juszczak10}
quote considerably larger values for $\Lambda_A$, in the range (1.20--1.35) GeV.

The polar-vector ($j^\mu_{\pm}$) and axial-vector  ($j^{\mu 5}_{\pm}$) components of the
charge-changing weak current are obtained, respectively, from $j^\mu_{\gamma,z}$ and $j^{\mu5}_{z}$ by replacing 
\begin{equation}
\tau_{i,z}/2 \longrightarrow \tau_{i,\pm}=(\tau_{i,x} \pm \tau_{i,y})/2 \ .
\label{eq:e12}
\end{equation}
However, in the case of $j^{\mu 5}_{\pm}$, in addition to the terms entering Eqs.~(\ref{eq:rho5})--(\ref{eq:j5}),  we also
retain the induced pseudoscalar contribution, given by
\begin{equation}
j^{\mu 5}_{\pm}(i;PS) = -{G_{PS}(Q^2)\over{2\,m\,m_\mu}}\,\tau_{i,\pm} \, q^\mu  \, {\bm \sigma}_i\cdot{\bf q} \, 
\, {\rm e}^{ i  {\bf q} \cdot {\bf r}_i } \ , 
\end{equation}
where the induced pseudoscalar form factor $G_{PS}$ is parametrized as 
\begin{equation}
G_{PS}(Q^2) = -{2\,m_\mu\,m \over m_\pi^2 + Q^2 }\, G_{A}(Q^2) \ .
\label{eq:gps}
\end{equation}
This form factor is not well known~\cite{Gorringe04}.  The parameterization
above is consistent with values extracted~\cite{Czarnecki07,Marcucci12} from precise
measurements of muon-capture rates on hydrogen~\cite{Andreev07} and $^3$He~\cite{Ackerbauer98},
as well as with the most recent  theoretical predictions based on chiral perturbation theory~\cite{Bernard94}. 
This contribution vanishes in NC-induced neutrino reactions.
\subsection{Two-body terms}
\label{sec:2b}
Two-body terms in the neutral and charge-changing weak
currents have been discussed in considerable detail in
Refs.~\cite{Carlson98,Marcucci00,Marcucci05} (and references therein).
We list the terms included in the present study---i.e., the subset of
those derived in the above references expected to give the dominant
two-body contributions to the processes of interest here---in the
following two sub-sections for clarity of presentation and future reference
in Sec.~\ref{sec:res}.  Unless stated otherwise, they are given in momentum
space, and configuration-space expressions follow from
\begin{equation}
O({\bf q})=
\int_{{\bf k}_i}\int_{{\bf K}_i} \int_{{\bf k}_j} \int_{{\bf K}_j} (2\pi)^3\,\delta({\bf k}_i+{\bf k}_j-{\bf q})\,
{\rm e}^{ i \,{\bf k}_i \cdot ({\bf r}^\prime_i+{\bf r}_i )/2}
{\rm e}^{ i \,{\bf K}_i \cdot ({\bf r}^\prime_i-{\bf r}_i )}
{\rm e}^{ i \,{\bf k}_j \cdot ({\bf r}^\prime_j+{\bf r}_j)/2 } 
{\rm e}^{ i \,{\bf K}_j \cdot ({\bf r}^\prime_j-{\bf r}_j )}
O({\bf k}_i,{\bf K}_i,{\bf k}_j,{\bf K}_j)\ , 
\label{jrs}
\end{equation}
where ${\bf k}_i={\bf p}_i^\prime-{\bf p}_i$ and ${\bf K}_i=({\bf p}^\prime_i+{\bf p}_i)/2$,
${\bf p}_i$ and ${\bf p}_i^\prime$ are the initial and final momenta of
nucleon $i$, and 
\begin{equation}
\int_{\bf p} \equiv \int \frac{{\rm d} {\bf p}}{(2\pi)^3} \ .
\end{equation}
These configuration-space operators
are used in the calculations reported below.
\subsubsection{Two-body vector terms}
\label{sec:2bv}
The two-body isovector current operator ${\bf j}_{\gamma,z}(ij)$ consists of
pseudoscalar- and vector-meson (referred to as $\pi$-like and $\rho$-like) exchange, and $\Delta$-excitation terms, 
\begin{equation}
{\bf j}_{\gamma,z}(ij)=\sum_{c=\pi,\,\rho,\, \Delta}\left[\, {\bf j}_{\gamma,z}(ij;c)+ i\rightleftharpoons j \,\right] \ .
\end{equation}
The $\pi$-like and $\rho$-like exchange currents read:
\begin{eqnarray}
\noalign{\medskip}
{\bf j}_{\gamma,z}(ij;\pi) 
&=&i\,G_{E}^{V}(Q^2)
   ({\bm\tau}_i \times {\bm\tau}_j)_z \, v_{\pi}(k_j)\, 
   \Bigg[  {\bm\sigma}_i 
 -{ {\bf k}_i - {\bf k}_j \over k_i^2 -k_j^2 }\,
   \left( {\bm\sigma}_i \cdot {\bf k}_i \right) \Bigg]\, {\bm\sigma}_j \cdot {\bf k}_j \ ,
  \label{eq:jps} \\
\noalign{\medskip}
  {\bf j}_{\gamma,z} (ij; \rho) 
   &=& - i\,G_{E}^{V}(Q^2) 
   ({\bm\tau}_i \times {\bm\tau}_j)_z 
 \Bigg[ 
   v_\rho(k_j) \,{\bm\sigma}_i \times ({\bm\sigma}_j \times {\bf k}_j) 
 +{ v_\rho(k_j) \over k_i^2 -k_j^2 } \Big [ 
       ({\bf k}_i-{\bf k}_j)({\bm\sigma}_i \times {\bf k}_i)\cdot
                            ({\bm\sigma}_j \times {\bf k}_j) \nonumber \\
\medskip
&&+ ({\bm\sigma}_i \times {\bf k}_i)\> 
  {\bm\sigma}_j \cdot ({\bf k}_i\times {\bf k}_j)
  +({\bm\sigma}_j \times {\bf k}_j)\> {\bm\sigma}_i \cdot 
  ({\bf k}_i\times {\bf k}_j) \Big] - v_{\rho S}(k_j)\,
  { {\bf k}_i - {\bf k}_j \over k_i^2 -k_j^2 }\, \Bigg ]
   \ , 
 \label{eq:jv}
\end{eqnarray}
where
\begin{equation}
   v_\pi(k)= v^{\sigma \tau}(k)-
2 \> v^{t\tau}(k) \ , \qquad
   v_\rho (k)=  v^{\sigma \tau}(k) + 
v^{t\tau}(k)  \ , \qquad
   v_{\rho S}(k)=v^{\tau}(k) \ ,
   \label{eq:vps}
   \end{equation}
   and
\begin{eqnarray}
v^\tau(k)&=&4\pi \int_0^\infty r^2 dr\, j_0(kr)v^\tau(r) \label{eq39}\>\>, \\
v^{\sigma \tau}(k)&=&\frac{4\pi}{k^2} \int_0^\infty r^2 dr \, 
\left [ j_0(kr)-1 \right ] v^{\sigma \tau}(r) \label{eq40}\>\>, \\
v^{t \tau}(k)&=&\frac{4\pi}{k^2} 
\int_0^\infty r^2 dr\, j_2(kr)v^{t\tau}(r) \label{eq41} \>\>.
\end{eqnarray}
Here $v^{\tau}(r)$, $v^{\sigma \tau}(r)$, $v^{t\tau}(r)$ are the
isospin-dependent central, spin-spin, and tensor components
of the two-nucleon interaction (the AV18 in the present study), and $j_l(kr)$ are
spherical Bessel functions.
The factor $j_0(kr)-1$ in the expression for $v^{\sigma \tau}(k)$
ensures that its volume integral vanishes.
In a one-boson-exchange (OBE) model, in which the isospin-dependent
central, spin-spin, and tensor interactions are due to $\pi$- and $\rho$-meson
exchange, the functions $v_\pi(k)$, $v_\rho(k)$, and $v_{\rho S}(k)$
simply reduce to
\begin{eqnarray}
v_\pi(k)&\longrightarrow&-\frac{f_\pi^2}{m_\pi^2}\frac{h^{\, 2}_\pi(k) }{k^2+m_\pi^2} \ , \\
v_\rho(k)&\longrightarrow &-\frac{g_{\rho}^2 \, (1+k_{\rho})^2}{4\, m^2}\, \frac{h^{\, 2}_\rho(k) }{k^2+m_\rho^2} \ , \\
v_{\rho S}(k)&\longrightarrow& g_{\rho}^2\, \frac{h^{\, 2}_\rho(k) }{k^2+m_\rho^2} \ ,
\end{eqnarray}
where $m_\pi$ and $m_\rho$ are the meson masses,
$f_\pi$, and $g_\rho$ and $\kappa_\rho$  are the pseudovector
$\pi NN$, and vector and tensor $\rho NN$ coupling constants, and the hadronic form factors are parameterized as
\begin{equation}
h_\alpha(k)=\frac{\Lambda_\alpha^2 -m_\alpha^2}{\Lambda_\alpha^2+k^2} \ , \qquad \alpha=\pi, \rho \ .
\label{eq:hff}
\end{equation}
While the AV18 interaction is not a OBE model, nevertheless
the effective propagators $v_\pi(k)$, $v_\rho(k)$, and $v_{\rho S}(k)$ projected
out of its $v^\tau(k)$, $v^{\sigma \tau}(k)$, and $v^{t\tau}(k)$ components
are quite similar to those listed above with cutoff masses in the range (1.0--1.5) GeV.
We note that the $\pi$-like and $\rho$-like currents with the $v_\pi(k)$, $v_\rho(k)$, and $v_{\rho S}(k)$
defined in Eq.~(\ref{eq:vps}) satisfy by construction the current
conservation relation with the AV18 $\tau$, $\sigma\tau$, and $t\tau$ interaction components
(for a discussion of the issue of current conservation in relation to the momentum-dependent terms
of the AV18, see Ref.~\cite{Marcucci05}).

The isovector $\Delta$-excitation current is written in configuration space as (for a derivation
based on a perturbative treatment of $\Delta$-isobar degrees of freedom in nuclear
wave functions, see Ref.~\cite{Carlson98})
\begin{equation} 
{\bf j}_{\gamma,z}(ij;\Delta)=- i \,\frac{G_{\gamma N \Delta}(Q^2)}{2\, m\, (m-m_\Delta)}\,
{\rm e}^{i {\bf q} \cdot {\bf r}_i}\, \Big[  v^\dagger_{\Delta N}(ij)\, {\bf q} \times {\bf S}_i \, T_{i,z}
+{\rm adjoint} \Big] \ ,
\label{eq:jdlt}
\end{equation}
where ${\bf S}$ and ${\bf T}$ are spin- and isospin-transition operators converting
a nucleon into a $\Delta$ isobar and satisfying the identity
\begin{equation}
{\bf S}^\dagger \cdot {\bf A} \,\, {\bf S}\cdot {\bf B} =\frac{2}{3} \, {\bf A}\cdot {\bf B}
-\frac{i}{3} {\bm \sigma} \cdot \left({\bf A} \times {\bf B}\right) \ ,
\end{equation}
$v_{\Delta N}(ij)$ is the $NN$ to $\Delta N$ transition potential,
\begin{equation}
v_{\Delta N}(ij)=\left[\, v^{\sigma \tau }_{\Delta N}(r_{ij})\, 
{\bf S}_i \cdot {\bm \sigma}_j +v^{t \tau}_{\Delta N}(r_{ij}) \, S^{\Delta N}_{ij} \, \right]
{\bf T}_i \cdot {\bm \tau}_j \ ,
\end{equation}
$S^{\Delta N}_{ij}$ is the tensor operator obtained by replacing ${\bm \sigma}_i$ with ${\bf S}_i$,
the regularized spin-spin and tensor radial functions $v^{\sigma \tau }_{\Delta N}(r)$
and $v^{t \tau }_{\Delta N}(r)$ are defined as
\begin{eqnarray}
v^{\sigma \tau}_{\Delta N}(r)&=&\frac {f_\pi  f_\pi^* }{4 \pi} \frac{ m_\pi}{3} \frac {{\rm e}^{-x}}{x}
 \left(1-{\rm e}^{-\lambda \, x^2} \right) \ , \\
v^{t \tau}_{\Delta N}(r)&=&\frac {f_\pi  f_\pi^*}{4 \pi} \frac{m_\pi}{3} \left (  1+\frac{3}{x}
+\frac{3}{x^2} \right ) \frac {{\rm e}^{-x}}{x}  \left(1-{\rm e}^{-\lambda \, x^2} \right)^2 \ ,
\end{eqnarray}
and $x=m_\pi r$, $f_\pi^*$ is the $\pi N\Delta$ coupling constant
($f_\pi^*=2.19 \, f_\pi$ from the width of the $\Delta$), and the parameter in the short-range
cutoff function is taken as $\lambda=4.29$ (from the AV18).   Finally, the $\gamma N \Delta$ electromagnetic transition
form factor $G_{\gamma N \Delta}$ is parameterized as
\begin{equation}
G_{\gamma N \Delta}(Q^2)= \frac{\mu_{\gamma N \Delta} }
{( 1+Q^2/\Lambda_{\Delta,1}^2 )^2
\sqrt{1+Q^2/\Lambda_{\Delta,2}^2} } \ ,
\end{equation}
where the transition magnetic moment $\mu_{\gamma N \Delta}$ is 3 n.m., as
obtained in an analysis of $\gamma N$ data
in the $\Delta$-resonance region~\cite{Carlson86}.  This
analysis also gives $\Lambda_{\Delta,1}$=0.84 GeV and
$\Lambda_{\Delta,2}$=1.2 GeV.  

The two-body isoscalar current operator ${\bf j}_{\gamma,S}(ij)$ considered in the
present study only includes the contribution associated with the $\rho \pi \gamma$
transition mechanism,
\begin{equation}
{\bf j}_{\gamma,S}(ij) = {\bf j}_{\gamma, S}(ij;\rho\pi) +  i \rightleftharpoons j \ ,
\end{equation}
where 
\begin{equation}
{\bf j}_{\gamma, S}(ij;\rho\pi)=-i\,G_{\rho\pi\gamma}(Q^2)\,
g_{\rho\pi\gamma}\,\frac{f_{\pi} }{m_\pi} \, \frac{ g_{\rho } }{m_\rho}\, \, {\bm \tau}_i \cdot {\bm \tau}_j \,
\frac{h_\rho(k_i)}{k_i^2+m_\rho^2}
\frac{h_\pi(k_j)}{k_j^2+m_\pi^2}
({\bf k}_i \times {\bf k}_j)\,  {\bm \sigma}_j \cdot {\bf k}_j \ ,
\label{eq:rpg_nr}
\end{equation}
The combination of coupling constants $g_{\rho\pi\gamma}\, f_\pi\, g_\rho$ is taken as
1.37, and the cutoff masses $\Lambda_\pi$ and $\Lambda_\rho$ as 0.75 GeV and
1.25 GeV, respectively, from a study of the deuteron magnetic form factor~\cite{Carlson91}. 
The $Q^2$ dependence of the electromagnetic transition form factor $G_{\rho\pi\gamma}(Q^2)$
is modeled by using vector-meson dominance,  
\begin{equation}
G_{\rho \pi \gamma}(Q^2)=\frac{1}{1+Q^2/m_\omega^2} \ ,
\end{equation}
where $m_\omega$ is the $\omega$-meson mass.

The two-body isovector and isoscalar electromagnetic charge operators $j^0_{\gamma,z}$ and $j^0_{\gamma,S}$
consist of terms associated with $\pi$-like and $\rho$-like exchanges
\begin{equation}
j^0_{\gamma,z/S}(ij)=\sum_{c=\pi,\,\rho}\left[\, j^0_{\gamma,z/S}(ij;c)+ i\rightleftharpoons j \,\right] \ ,
\end{equation}
where
\begin{eqnarray}
j^0_{\gamma,z}(ij;\pi)&=&\frac{ F_1^V(Q^2)} {2\, m} \,
\tau_{z,j} \,
v_\pi(k_j) \, ({\bm \sigma}_i \cdot {\bf q})\, (  {\bm \sigma}_j \cdot {\bf k}_j)  \ ,
\label{eq68a}  \\
j^0_{\gamma,z} (ij;\rho)&=&\frac {F_1^V(Q^2)}{2\, m} \, 
\tau_{z,j} \,
 v_\rho(k_j)\, ( {\bm \sigma}_i \times {\bf q}) \cdot ({\bm \sigma}_j \times 
{\bf k}_j) \ ,
\label{eq69a} 
\end{eqnarray}
and 
\begin{eqnarray}
j^0_{\gamma,S}(ij;\pi)&=&\frac{F_1^S(Q^2)} {2\, m} \, {\bm \tau}_i \cdot {\bm \tau}_j  \,
v_\pi(k_j) \, ({\bm \sigma}_i \cdot {\bf q})\, (  {\bm \sigma}_j \cdot {\bf k}_j) \ ,
\label{eq68} \\
j^0_{\gamma,S} (ij;\rho)&=&\frac {F_1^S(Q^2)}{2\, m} \,
{\bm \tau}_i \cdot  {\bm \tau}_j  \,
 v_\rho(k_j)\, ( {\bm \sigma}_i \times {\bf q}) \cdot ({\bm \sigma}_j \times 
{\bf k}_j) \ ,
\label{eq69} 
\end{eqnarray}
with $v_\pi(k)$ and $v_\rho(k)$ as defined in Eqs.~(\ref{eq:vps}).  The nucleon
electromagnetic Dirac and Pauli form factors $F^{S/V}_1$ and $F^{S/V}_2$are obtained from
\begin{eqnarray}
\label{eq:f1ff}
F_1^{S/V}\!(Q^2)&=&\frac{G_E^{S/V}\!(Q^2) +\eta \, G_M^{S/V}\!(Q^2)}{1+\eta} \ , \\
F_2^{S/V}\!(Q^2)&=&\frac{G_M^{S/V}\!(Q^2)-G_E^{S/V}\!(Q^2)}{1+\eta}  \ ,
\label{eq:f2ff}
\end{eqnarray}
with $\eta=Q^2/(4\,m^2)$.  

The polar-vector components $j^\mu_{\pm}$ of the charge-changing
weak current $j^\mu_{CC}$
are obtained from $j^\mu_{\gamma,z}$ via CVC, which implies the replacements
\begin{equation}
({\bm \tau}_i \times {\bm \tau}_j)_z  \longrightarrow ({\bm \tau}_i \times {\bm \tau}_j)_\pm=
({\bm \tau}_i \times {\bm \tau}_j)_x \pm i \, ({\bm \tau}_i \times {\bm \tau}_j)_y
\label{eq:e44}
\end{equation}
 in Eqs.~(\ref{eq:jps})--(\ref{eq:jv}), 
 \begin{equation}
 T_{i,z}/2  \longrightarrow T_{i, \pm} =( T_{i,x}\pm i\, T_{i,y} )/2
 \label{eq:e45}
 \end{equation}
in Eq.~(\ref{eq:jdlt}), and the replacement~(\ref{eq:e12}) in Eqs.~(\ref{eq68a})--(\ref{eq69a}).
Only the transverse components (perpendicular to ${\bf q}$) of the
vector part of the NC and CC are explicitly included in the calculations to follow.  Their
longitudinal components have already been effectively accounted for by the
replacement in Eq.~(\ref{eq:jlong}) (and the similar one for the isovector terms).  Lastly,
we note that in the study of Ref.~\cite{Nakamura02} the $\rho$-meson exchange and $\rho\pi$
transition contributions to the two-body vector current, and $\pi$- and $\rho$-exchange 
contributions to the two-body vector charge  have been neglected.  Furthermore, the $\pi$-exchange
and $\Delta$ excitation currents are regularized by introducing a monopole form factor ($\Lambda_\pi=4.8$
fm$^{-1}$), which naturally leads to a different short-range behavior of these currents than
obtained here. 
 \subsubsection{Two-body axial terms}
 \label{sec:2ba}
\begin{table}[bthp]
\caption{Contributions to the GT matrix element
in tritium $\beta$-decay.  The one-body (1-b) NR and RC contributions
are, respectively, from the leading and $1/m^2$ terms in Eq.~(\protect\ref{eq:j5});
the two-body (2-b) contributions are from Eqs.~(\protect\ref{eq:a2pi})--(\protect\ref{eq:jdlta}).
Set I (II) corresponds to the
cutoff choices $\Lambda_\pi=\Lambda_\rho=1.2$ GeV ($\Lambda_\pi=1.72$ GeV
and $\Lambda_\rho=1.31$ GeV).  The $N$ to $\Delta$ axial
coupling constant $g_A^*$ for each set is obtained by fitting the experimental value
of the GT matrix element, given by $0.955 \pm 0.002$~\protect\cite{Marcucci12}.}
\begin{tabular}{c|c|c}
\hline\hline
  &  Set I & Set II \\
\hline\hline
1-b (NR)    & +0.9213   &+0.9213\\
1-b (RC)    & --0.0085     &--0.0085  \\
\hline
2-b ($\pi$)    & +0.0078   &+0.0123 \\
2-b ($\rho$)  &  --0.0042 & --0.0055 \\
2-b ($\rho\pi$)     & +0.0123  &+0.0196 \\
2-b ($\Delta$)   & +0.0263    & +0.0159\\
 \hline
$g_A^*/g_A$  &  0.614    & 0.371  \\
\hline\hline
\end{tabular}
\label{tab:gas}
\end{table}

The axial parts of the neutral and charge-changing weak current operators
consist  of contributions associated with $\pi$- and $\rho$-meson exchanges,
the axial $\rho \pi$ transition mechanism, and a $\Delta$ excitation term
\begin{equation}
{\bf j}^5_{a}(ij)=\sum_{c=\pi,\,\rho,\, \rho\pi,\, \Delta}\left[\, {\bf j}^5_{a}(ij;c)+ i\rightleftharpoons j \,\right] \ ,
\end{equation}
where the isospin component $a$ is either $z$ for NC or $\pm$ for CC.
The $\pi$- and $\rho$-meson exchange and $\rho\pi$ transition axial currents read, respectively,
\begin{equation}
{\bf j}^5_z(ij;\pi)=\frac{G_A(Q^2)}{2\, m} \, \frac{f^2_\pi}{m^2_\pi} \frac{h^2_\pi(k_j)}{k_j^2+m_\pi^2}
 \Big[ ({\bm \tau}_i \times {\bm \tau}_j)_z \,
 {\bm \sigma}_i \times {\bf k}_j  
 -\tau_{j,z} \,  \left ( {\bf q}  + 2\, i\, {\bm \sigma}_i
 \times {\bf K}_i \right ) \Big] \,  {\bm \sigma}_j \cdot {\bf k}_j\ ,
\label{eq:a2pi}
\end{equation}
\begin{eqnarray}
{\bf j}^5_z(ij;\rho)&=& -\frac{G_A(Q^2)}{2\, m}\,\frac{g_{\rho}^2 \, 
(1+k_{\rho})^2}{4\, m^2}\, \frac{h^{\, 2}_\rho(k_j) }{k_j^2+m_\rho^2} \Bigg[
({\bm \tau}_i \times {\bm \tau}_j)_z \,
\Big [\, {\bf q} \,\,{\bm \sigma}_i \cdot ({\bm \sigma}_j
\times {\bf k}_j)+2\, i\, ({\bm \sigma}_j \times {\bf k}_j) 
\times {\bf K}_i \nonumber \\
&&-  \left[{\bm \sigma}_i \times ({\bm \sigma}_j \times {\bf k}_j)  \right] \times {\bf k}_j\Big ] 
+\tau_{j,z}\, 
\Big [ ({\bm \sigma}_j \times  {\bf k}_j) \times {\bf k}_j -2\, i\,
\left[ {\bm \sigma}_i \times ({\bm \sigma}_j \times {\bf k}_j ) \right]
\times {\bf K}_i \Big ]  \Bigg]\ ,
\label{eq:a2ro}
\end{eqnarray}
\begin{eqnarray}
{\bf j}^5_z(ij;\rho\pi) &=&
 - {G_A(Q^2)\over m} \, g_\rho^2 \,  
{h_\rho(k_i) \over k_i^2 + m_\rho^2} { h_\pi(k_j) \over k_j^2 + m_\pi^2 } \,
({\bm \tau}_i \times {\bm \tau}_j)_z \,  
 \Big [ (1 + \kappa_\rho )\, {\bm \sigma}_i \times {\bf k}_i  
- 2\, i\, {\bf K}_i \Big ]  {\bm \sigma}_j \cdot {\bf k}_j   \ ,
\label{eq:a2rp}
\end{eqnarray}
while the $\Delta$ excitation axial current is obtained from~\cite{Carlson98}
\begin{equation}
{\bf j}^5_z(ij;\Delta)=-\frac{G^*_A(Q^2)}{2\,(m-m_\Delta)}\,
{\rm e}^{i {\bf q} \cdot {\bf r}_i}\, \Big[  v^\dagger_{\Delta N}(ij)\,{\bf S}_i \, T_{i,z}
+{\rm adjoint} \Big] \ ,
\label{eq:jdlta}
\end{equation}
where the (unknown) $N$ to $\Delta$ axial form factor is parameterized as
\begin{equation}
 G^*_A(Q^2) = \frac{g^*_A}{\left(1+Q^2/\Lambda_A^2\right)^2} \ .
\end{equation}
The charge-changing axial currents follow by replacing
the isospin operators as in Eqs.~(\ref{eq:e12}) and~(\ref{eq:e44})--(\ref{eq:e45}).
The values for the $\pi$- and $\rho$-meson coupling constants are taken from
the CD-Bonn one-boson-exchange model~\cite{Machleidt01}, $f_\pi^2/(4\, \pi)=0.075$,
$g_\rho^2/(4\, \pi)=0.84$, and $\kappa_\rho=6.1$, while two different sets
of cutoff masses $\Lambda_\pi$ and $\Lambda_\rho$ are used in the
present work: $\Lambda_\pi=\Lambda_\rho=1.2$ GeV (Set I) in line with
the cutoff masses extracted from the $\pi$-like and $\rho$-like
exchanges associated with the AV18 model;
$\Lambda_\pi=1.72$ GeV and $\Lambda_\rho=1.31$ GeV (Set II) from
the CD-Bonn model.  In the $N$ to $\Delta$ axial current, the $Q^2$ dependence
of the form factor is taken to be the same as that of the nucleon; however, the value
for the transition axial coupling constant $g_A^*$ is determined by fitting the
Gamow-Teller (GT) matrix element of tritium $\beta$-decay~\cite{Marcucci12}
in a calculation based on trinucleon wave functions corresponding to the AV18/UIX Hamiltonian
and the present model for the axial current.  The values corresponding
to Set I and II of cutoff masses are listed in Table~\ref{tab:gas}.

Finally, in the present study the axial charge operator is taken to include
only the pion-exchange term, whose structure and strength are determined by
soft-pion theorem and current algebra arguments~\cite{Kubodera78}
\begin{equation}
j^{05}_{a}(ij)= j^{05}_{a}(ij;\pi)+ i\rightleftharpoons j \ ,
\end{equation}
where
\begin{equation}
j^{05}_a(ij;\pi)=
-i\,\frac{G_A(Q^2)}{4 \,F_\pi^2}\,\frac{h_\pi^2(k_i) }{ k_i^2+ m_\pi^2 }\, 
({\bm \tau}_i \times{\bm \tau}_j)_a \,
{\bm \sigma}_i \cdot{\bf k}_i \ ,
\label{eq:rho2pi}
\end{equation}
$F_\pi$ is pion decay amplitude ($F_\pi\,$= 93 MeV), and
the $Q^2$ dependence of the associated form factor is assumed to be the
same as in the nucleon.  We conclude by noting that the model described above
for the two-body axial charge and current operators is essentially the same
of that used in Ref.~\cite{Nakamura02}, apart from differences in the values
of the cutoff masses for the hadronic form factors of the meson exchange terms, and
a different treatment of the $\Delta$ excitation current.  However, it is important
to stress that both here and in Ref.~\cite{Nakamura02} the two-body axial
currents are constrained to reproduce the experimental tritium $\beta$-decay rate.
\section{Calculation}
\label{sec:calc}
The two-body scattering- and bound-state problems are solved in momentum space
with the methods discussed in Ref.~\cite{Schiavilla04}, which facilitates calculations
with a non-local potential such as CD Bonn.  We briefly summarize them
in the next two sub-sections for clarity.  In the last sub-section we discuss
the calculation of the weak current matrix elements, response functions,
and cross sections.   

\subsection{The scattering-state problem in momentum space}

In the case of scattering (setting aside the treatment of the Coulomb
interaction for the time being), we solve for the $K$-matrix in channel
$JST$ (hereafter, $L$ is the relative orbital angular momentum, $S$ and $T$ are
the total spin and isospin, and $J$ is the total angular momentum, and $(-1)^{L+S+T}=-1$)
\begin{equation}
K^{JST}_{L^\prime L}(p^\prime ;p)=
v^{JST}_{L^\prime L}(p^\prime ;p)
+\frac{4\mu}{\pi} \int_0^\infty {\rm d}k\,k^2 \sum_{L^{\prime\prime}}
v^{JST}_{L^\prime L^{\prime\prime}}(p^\prime;k)
\frac{\cal P}{p^2-k^2}K^{JST}_{L^{\prime\prime} L}(k;p) \>\>,
\label{eq:kma}
\end{equation}
where $\mu$ is the two-nucleon reduced mass,
${\cal P}$ denotes a principal-value integration,
and $v^{JST}_{L^\prime L}(p^\prime\!,p)$ are  the $p$-space
matrix elements of the potential, projected in channel $JST$~\cite{Schiavilla04}.
We should note the presence of the somewhat unconventional
phase factor $i^{L-L^\prime}$ included in the matrix elements
$v^{JST}_{L^\prime L}(p^\prime ;p)$~\cite{Schiavilla04}, which
makes the states used here differ by a factor $i^L$ from those
usually adopted in nucleon-nucleon scattering analyses.
The integral equations~(\ref{eq:kma}) are discretized, and the resulting
systems of linear equations are solved by direct numerical
inversion.  The principal-value integration is eliminated by
a standard subtraction technique~\cite{Gloeckle83}.  Phase shifts
in channel $JST$ are easily obtained from the on-shell $S$-matrix
related to the (on-shell) $K$-matrix by
\begin{equation}
S^{JST}(p)=\left[ 1+2\, i \, \mu\, p\, K^{JST}(p;p) \right]^{-1}
\left[ 1-2\, i \, \mu \, p\, K^{JST}(p;p) \right] \ ,
\label{eq:skma}
\end{equation}
while $r$-space wave functions follow from 
\begin{equation}
z^{JST}_{L^\prime L}(r;p)=\Bigg[ j(pr)
+\frac{4\mu}{\pi}\int_0^\infty {\rm d}k\, k^2\,
j(kr) \frac{\cal P} {p^2-k^2}
K^{JST}(k;p)\Bigg]_{L^\prime L^{\prime\prime}} \, \Bigg[ 1+ 2\, i\, \mu\, p\, K^{JST}(p;p)\Bigg]^{-1}_{L^{\prime\prime} L} \ ,
\label{eq:psirk}
\end{equation}
where the matrix $[j(pr)]_{L^\prime L}\equiv \delta_{L^\prime L}\, j_{L}(pr)$
has been introduced for convenience.
The (complex) radial wave functions $z^{JST}_{L^\prime L}(r)$
behave in the asymptotic region $r \rightarrow \infty$ as
\begin{equation}
z^{JST}_{L^\prime L}(r;p)\simeq \frac{1}{2} \Big[
\delta_{L^\prime  L} h^{(2)}_{L}(pr)
+h^{(1)}_{L^\prime}(pr) S^{JST}_{L^\prime L}(p) \Big] \ ,
\label{eq:asy}
\end{equation}
where the functions $h_L^{(1,2)}(pr)$ are defined in terms of the regular
and irregular ($n_L$) spherical Bessel functions as
\begin{equation}
h^{(1,2)}_L(y)=j_L(y) \pm i\, n_L(y) \ .
\end{equation}
In the calculation of the response functions that follows, scattering wave 
functions with incoming-wave boundary condition $(-)$ are required.  These
are written as
\begin{eqnarray}
\psi^{(-)}_{SM_S,TM_T}({\bf r};{\bf p})&=&
4\pi\sqrt{2} \!\!\!\sum_{JM_J, J \le J_{\rm max}} \sum_{L L^\prime}\,
i^{L^\prime} Z_{LSM_S}^{JM_J *}(\hat{\bf p})\, \Big[ 
 z^{JST *}_{L^\prime L}(r;p) -\delta_{L^\prime  L}\, j_L(pr)\Big]\,
{\cal Y}_{L^\prime S  J}^{M_J}(\hat{\bf r})\, \eta^{T}_{M_T} \nonumber \\
&&+\frac{1}{\sqrt{2}} \left[ {\rm e}^{ i\, {\bf p} \cdot {\bf r} }
-(-)^{S+T} {\rm e}^{-i\, {\bf p} \cdot {\bf r} } \right]
\chi^S_{M_S}\, \eta^T_{M_T} \ ,
\label{eq:psipw}
\end{eqnarray}
where $\chi^S_{M_S}$ and $\eta^T_{M_T}$ are two-nucleon
spin and isospin states, respectively, ${\cal Y}_{L S  J}^{M_J}$ are standard spin-angle functions,
\begin{equation}
Z_{LSM_S}^{JM_J}(\hat{\bf p})\equiv
\sum_{M_L} \langle LM_L,SM_S\mid JM_J\rangle
\, Y_{LM_L}(\hat {\bf p}) \ ,
\end{equation}
and $ \langle LM_L,SM_S\mid JM_J\rangle$ are Clebsch-Gordan coefficients.  Note that 
the wave function in Eq.~(\ref{eq:psipw}) retains interaction effects only in channels with
$J \le J_{\rm max}$, and reduces to plane waves for $J > J_{\rm max}$. 

When the Coulomb interaction is present, we use the
method developed originally in Ref.~\cite{Vincent74}, which allows
us to solve the $pp$ scattering problem in momentum space~\cite{Carlson02}.
It consists essentially in separating the potential into short-
and long-range parts $v_S$ and $v_L$, where $v_L$ only includes the Coulomb
potential $v_C$ and $v_S$ includes, in addition to $v_C$, the nuclear potential $v$. 
Then the standard momentum-space technique outlined earlier can be used to
solve the problem with $v_S$, and the corresponding radial wave functions
behave as 
\begin{equation}
z^{JS1}_{S;L^\prime L}(r;p)\simeq \frac{a_L}{2} \Big[
\delta_{L^\prime  L} h^{(2)}_{L}(pr)
+h^{(1)}_{L^\prime}(pr) S^{JS1}_{S;L^\prime L}(p) \Big] \ ,
\label{eq:asys}
\end{equation}
where $S^{JS1}_S$ is the $S$-matrix in this case (with $T=1$), and the
$a_L$ are normalization constants.  The wave functions $z^{JS1}_{S; L^\prime L}$
should match smoothly those relative to $v_S+v_L$, which behave
asymptotically as
\begin{equation}
z^{JS1}_{L^\prime L}(r;p)\simeq \frac{1}{2} \Big[
\delta_{L^\prime  L} {\overline h}^{(2)}_{L}(\xi,pr)
+\overline{h}^{(1)}_{L^\prime}(\xi, pr)\, S^{JS1}_{L^\prime L}(p) \Big] \ ,
\label{eq:asyc}
\end{equation}
where
\begin{equation}
\overline{h}^{(1,2)}_L(\xi,y)=\left[F_L(\xi; y) \mp G_L(\xi;y)\right]/y \ , \qquad \xi=\alpha \, \mu/p \ ,
\end{equation}
and $F_L$ and $G_L$ are the regular and irregular Coulomb functions.
Carrying out the matching for the functions and their
first derivatives leads to a relation between $S^{JS1}_S$ and $S^{JS1}$ and a
determination of the normalization constants~\cite{Carlson02}.  Finally,
$pp$ scattering wave functions with incoming-wave boundary
conditions are written as in Eq.~(\ref{eq:psipw}) with $T,M_T=1,1$ and the replacement 
\begin{equation}
z^{JS1 *}_{L^\prime L}(r;p) \longrightarrow {\rm e}^{-i\, \sigma_L}\, z^{JS1 *}_{L^\prime L}(r;p) \ ,
\end{equation}
where $\sigma_L$ is the Coulomb phase shift,
\begin{equation}
\sigma_L={\rm arg}\left[ \Gamma(L+1+i\, \xi)\right] \ .
\end{equation}
Hence, Coulomb interaction effects are retained only in channels with
$J \le J_{\rm max}$, and are ignored in those with $J > J_{\rm max}$.
\subsection{The bound-state problem in momentum space}
The deuteron wave function is written in $r$-space as
\begin{equation}
\psi_{M}({\bf r})=\sum_{L=0,2} i^L  \, u_{L}(r) \, 
{\cal Y}_{L11}^{M}(\hat{\bf r})\, \eta^0_0 \ ,
\label{eq:dw}
\end{equation}
and the radial wave functions $u_L(r)$ ($L=0,2$) follow from 
\begin{equation}
u_L(r)=\frac{2}{\pi} \int_0^\infty {\rm d}p\, p^2\, j_L(pr)\, \overline{u}_{L}(p) \ .
\end{equation}
 The $p$-space wave functions $\overline{u}_L(p)$ are obtained
from solution of the homogeneous integral equations
\begin{equation}
\overline{u}_{L}(p)=\frac{1}{E_d-p^2/(2\mu)} \frac{2}{\pi}
\int_0^\infty {\rm d}k\, k^2 \sum_{L^\prime=0,2}
v^{110}_{L L^\prime}(p;k)\, \overline{u}_{L^\prime}(k) \ .
\end{equation}
Here, $v^{110}_{L^\prime L}$ is the nuclear potential in the
$JST=110$ channel, and $E_d$ is the deuteron energy ($E_d=-2.225$ MeV).
We again note the unconventional phase $i^L$ in Eq.~(\ref{eq:dw}).
\subsection{Matrix elements, response functions, and cross sections}

The deuteron wave function in Eq.~(\ref{eq:dw}) is written, for each
spatial configuration ${\bf r}$, as a vector in the spin-isospin
space of the two nucleons,
\begin{equation}
\psi_{M}({\bf r})=\sum_{n=1}^8 \psi_{M}^{(n)}({\bf r}) \mid\! n\rangle \ ,
\end{equation}
where $\mid\!\! n\rangle = (p\!\uparrow)_1\, (n\!\uparrow)_2, (n\!\uparrow)_1\, (p\!\uparrow)_2, \dots,
(n\!\downarrow)_1\, (p\!\downarrow)_2$ and $\psi_{M}^{(n)}$ are the components
of $\psi_{M}$ in this basis.  In the NC-induced processes, the scattering wave function
in Eq.~(\ref{eq:psipw}) is expanded on the same basis; however, in the CC-induced
processes the $pp$ or $nn$ scattering wave functions are expanded on the
(spin only) basis $\mid \! m\rangle = \uparrow   \uparrow \ , \downarrow  \uparrow\ , 
\uparrow \downarrow $, and $\downarrow  \downarrow$ for $pp$ or
$nn$.  Matrix elements of the weak current operators are
written schematically as
\begin{equation}
\langle\, f\mid O \mid d, M\rangle = \int{\rm d}{\bf r}\sum_{n^\prime,n}
\psi_f^{(n^\prime)\,*}({\bf r})\, O_{n^\prime,n}({\bf r}) \, \psi_M^{(n)}({\bf r}) \ ,
\end{equation}
where the momentum- and energy-transfer dependence is understood.
The spin-isospin algebra is performed exactly with techniques similar to
those developed in Ref.~\cite{Schiavilla89}, while the ${\bf r}$-space
integrations are carried out efficiently by Gaussian quadratures.  Note 
that no multipole expansion of the transition operators is required.
When momentum operators are present, they are taken to act on the right (deuteron)
wave function.  For example, the one-body axial charge operator is
written as
\begin{equation}
O({\bf r})\, \psi_M({\bf r}) \longrightarrow  -\frac{G_A(Q^2)}{4\,m}\left [ {\rm e}^{i \, {\bf q}\cdot {\bf r}/2} \,  
{\bm \sigma}_1 \cdot \left(-2\, i {\bm \nabla}+{\bf q}\right)\, \tau_{1,z} +  1 \rightleftharpoons 2\right]
\psi_M({\bf r}) \ ,
\label{eq:e16}
\end{equation}
and the derivatives are evaluated numerically as
\begin{equation}
\nabla_\alpha \psi_M({\bf r}) \simeq \frac{\psi_M({\bf r}+\delta \, \hat{\bf e}_\alpha)-
\psi_M({\bf r}-\delta\, \hat{\bf e}_\alpha)}{2\, \delta} \ ,
\end{equation}
where $\hat{\bf e}_\alpha$ is a unit vector in the $\alpha$-direction, and $\delta$ is a small increment.
Once the matrix elements have been computed, response functions
are evaluated (in the lab frame) via
\begin{equation}
R_{ab}(q,\omega)= \frac{1}{3} \sum_M \sum_{S M_S,T} \int \frac{{\rm d} {\bf p}}{(2\pi)^3} \frac{1}{2} \,
 \delta( \omega+m_d-E_+-E_-)\, f^{SM_S,TM_T;M}_{ab}({\bf q},{\bf p})  \ , 
\label{eq:e1}
\end{equation}
with
\begin{equation}
f^{SM_S,TM_T;M}_{ab}({\bf q},{\bf p})= \langle {\bf q},{\bf p}; SM_S,TM_T |\, O_a({\bf q},\omega )\, |d,M\rangle
\langle {\bf q},{\bf p} ; SM_S,TM_T |\, O_b({\bf q};\omega)\, |d,M\rangle^*\ ,
\end{equation}
where $ \mid\!\! {\bf q},{\bf p}; SM_S,TM_T\rangle$ represents the final two-nucleon scattering
state with total momentum ${\bf q}$ and relative momentum ${\bf p}$, $m_d$ is the deuteron rest mass,
and $E_\pm$ are the nucleons' energies in the final state,
\begin{equation}
E_\pm =\sqrt{({\bf q}/2\pm {\bf p})^2+m^2}\ .
\end{equation} 
The factor 1/2 in Eq.~(\ref{eq:e1}) is to avoid double-counting the contribution of the final states (which are
anti-symmetrized), and the pair isospin $T$ assumes the values $T=0,1$ with $M_T=0$ for NC
processes, and $T=1$ with $M_T=1$ or $-1$ for CC processes.  The $\delta$-function is integrated out,
and 
\begin{equation}
R_{ab}(q,\omega)=\frac{1}{24\, \pi^2} \sum_M \sum_{S M_S,T} \int^{+1}_{-1}{\rm d} x\, 
p^2\, \Bigg | \frac{p+x \, q/2}{E_+} + \frac{p-x \, q/2}{E_-}\Bigg |^{-1}
f^{SM_S,TM_T;M}_{ab}(q,p,x) \ ,
\end{equation}
where $x=\hat{\bf q} \cdot \hat {\bf p}$, and the magnitude $p$ of the relative momentum
is fixed by energy conservation.  This magnitude depends on $q$, $\omega$, and $x$.  However,
in order to reduce the computational effort, the scattering states entering the product of matrix elements
$f_{ab}$ are obtained at the energy 
\begin{equation}
4\, (\overline{p}^{\, 2}+m^2)= (\omega+m_d)^2-q^2\ , 
\label{eq:pbar}
\end{equation}
which only depends on $q$ and $\omega$.  Lastly, Gauss points ($\sim 50$) are used to
perform the $x$-integration accurately.  Extensive and independent tests of the computer programs have been
completed successfully.

Total cross sections are obtained by direct integration of Eq.~(\ref{eq:xsw}) by evaluating
the differential cross sections on a grid of Gauss points in $\epsilon^\prime$ (the
lepton final energy) and $\theta$ (its scattering angle).  There are kinematical constraints on the
allowed values for $\epsilon^\prime$ and $\theta$, which follow from the requirement
$\overline{p}^{\, 2} \ge 0$:
\begin{equation}
\epsilon\, \sqrt{\epsilon^{\prime\, 2}- m_l^2} \, {\rm cos}\, \theta
\ge (\epsilon+m_d)\,(\epsilon^\prime -\overline{\epsilon}) \ , \qquad
\overline{\epsilon}=\frac{ m_d(\epsilon-\epsilon_{\rm th})+m_l(m_l+2\, m)}{\epsilon+m_d} \ ,
\end{equation}
where $\epsilon_{\rm th}$ is the threshold energy for the initial neutrino ($\epsilon > \epsilon_{\rm th}$),
\begin{equation}
\epsilon_{\rm th}=\frac{(m_l+2\, m)^2-m_d^2}{2\, m_d} \ ,
\end{equation}
$m_l$ is the rest mass of the final lepton ($m_l=0$ in the NC case),
and $m=(m_p+m_n)/2$ for NC reactions or $m=m_p$ ($m_n$)
for charge-raising (charge-lowering) reactions.  These kinematical constraints
imply:
\begin{eqnarray}
&&m_l \le \epsilon^\prime \le \epsilon^\prime_- \,\,\,\,\,\, {\rm for} \,\,\,\,\,\, -1 \le {\rm cos}\, \theta \le 0 \ , \\
&&m_l \le \epsilon^\prime \le \epsilon^\prime_+ \,\,\,\,\,\, {\rm for} \,\,\,\,\,\,\,\,\,\,\,\,\,\, 0 \le {\rm cos}\, \theta \le 1 \ ,
\end{eqnarray}
where the limits $\epsilon^\prime_{\pm}$ are defined as
\begin{equation}
\epsilon^\prime_{\pm}=\frac{ \overline{\epsilon} \pm \sqrt{ \overline{\epsilon}^{\, 2}
-\left( 1-\beta^{\, 2}\,{\rm cos}^2\, \theta\right)  \left( \overline{\epsilon}^{\, 2} +m_l^2\, \beta^{\, 2}\,{\rm cos}^2\, \theta \right)}}
{1-\beta^{\, 2}\, {\rm cos}^2\, \theta} \ , \qquad \beta =\frac{1}{1+m_d/\epsilon} \ .
\end{equation}
In the case of NC reactions ($m_l=0$), they are simply given by
\begin{equation}
0 \le \epsilon^\prime \le \frac{\overline{\epsilon}}{1-\beta\, {\rm cos}\, \theta}
 \,\,\,\,\,\, {\rm for} \,\,\,\,\,\, -1 \le {\rm cos}\, \theta \le 1\ .
\end{equation}
\section{Results}
\label{sec:res}
Cross section values obtained with the AV18 interaction and the one- and two-body terms
in the electroweak current discussed in Sec.~\ref{sec:cnts} are listed in Tables~\ref{tab:gnu}--\ref{tab:cc}
for initial neutrino energies in the range (5--1000) MeV.  The two-body axial currents are those
corresponding to Set I in Table~\ref{tab:gas}.  The two-nucleon ($NN$) scattering states are written
as in Eq.~(\ref{eq:psipw}): they include interaction effects in channels with $J \le J_{\rm max}=5$
and reduce to spherical Bessel functions (i.e., plane waves) for $J> J_{\rm max}$.  The relative
kinetic energy $T=2\,(\overline{p}^2+m^2)^{1/2}-2\, m$, where $\overline{p}$ is defined in Eq.~(\ref{eq:pbar}), 
changes over a wide range as the initial neutrino energy increases up to 1 GeV and the final
lepton energy and scattering angle vary over the allowed kinematical regions: at $\epsilon=50$ MeV,
$0  \lesssim T \lesssim 48 $ MeV; at $\epsilon=500$ MeV,  $0  \lesssim T \lesssim 445$ MeV;
and at $\epsilon=1000$ MeV,  $0  \lesssim T \lesssim 819$ MeV.
These relative energies (at the larger values of $\epsilon$) are well beyond the range
of applicability of all modern realistic interactions, which are typically constrained to fit
$NN$ scattering data up to pion production threshold ($T\simeq 150$ MeV).  This is
also the case for the AV18 of course, although it is known~\cite{Wiringa12} that this interaction
reproduces quite well phase shifts (at least in those channels where inelasticities
are small) up to $T \lesssim 300$ MeV.
\begin{table}[bth]
\caption{Total cross sections in cm$^2$ for the NC- and CC-induced processes on the deuteron
as function of the initial neutrino energy $\epsilon$, obtained with the AV18
potential and the inclusion of one- and two-body terms in the weak current. The number in parentheses, ``$-x$", denotes
10$^{-x}$; for instance an entry like 9.561(--44) stands for 9.561$\times 10^{-44}$ cm$^2$. } 
\begin{tabular}{c|c|c|c|c|}
$\epsilon$ (MeV) & $\nu_l$-NC & $\overline{\nu}_l$-NC & $\nu_e$-CC& $\overline{\nu}_e$-CC \\
\hline
5   &  9.561(--44)     & 9.363(--44) &   3.427(--43)  &   2.831(--44)   \\
10 &  1.104(--42)     & 1.053(--42) &  2.680(--42)   &   1.242(--42)  \\
20 &  6.965(--42)     & 6.285(--42) &  1.547(--41)   &   9.562(--42) \\
30 &  1.833(--41)     & 1.568(--41) &  4.058(--41)   &   2.508(--41)  \\
40 &  3.555(--41)     & 2.885(--41) &  7.995(--41)   &    4.685(--41) \\
50 &  5.892(--41)     & 4.546(--41) &   1.348(--40)  &   7.403(--41)   \\
60 &  8.839(--41)     & 6.495(--41) &    2.338(--40) &   1.057(--40) \\
70 &  1.240(--40)     & 8.699(--41) &   2.949(--40)   &  1.409(--40) \\
80 &  1.657(--40)     & 1.111(--40)  &  4.036(--40)    &  1.790(--40) \\
90 &   2.131(--40)    & 1.369(--40)  &   5.320(--40)   &  2.191(--40)\\
100 &  2.657(--40)     & 1.640(--40) &    6.631(--40) &  2.606(--40) \\
\hline
\end{tabular}
\label{tab:gnu}
\end{table}

\begin{figure}[bth]
 \centering
\includegraphics[width=3in]{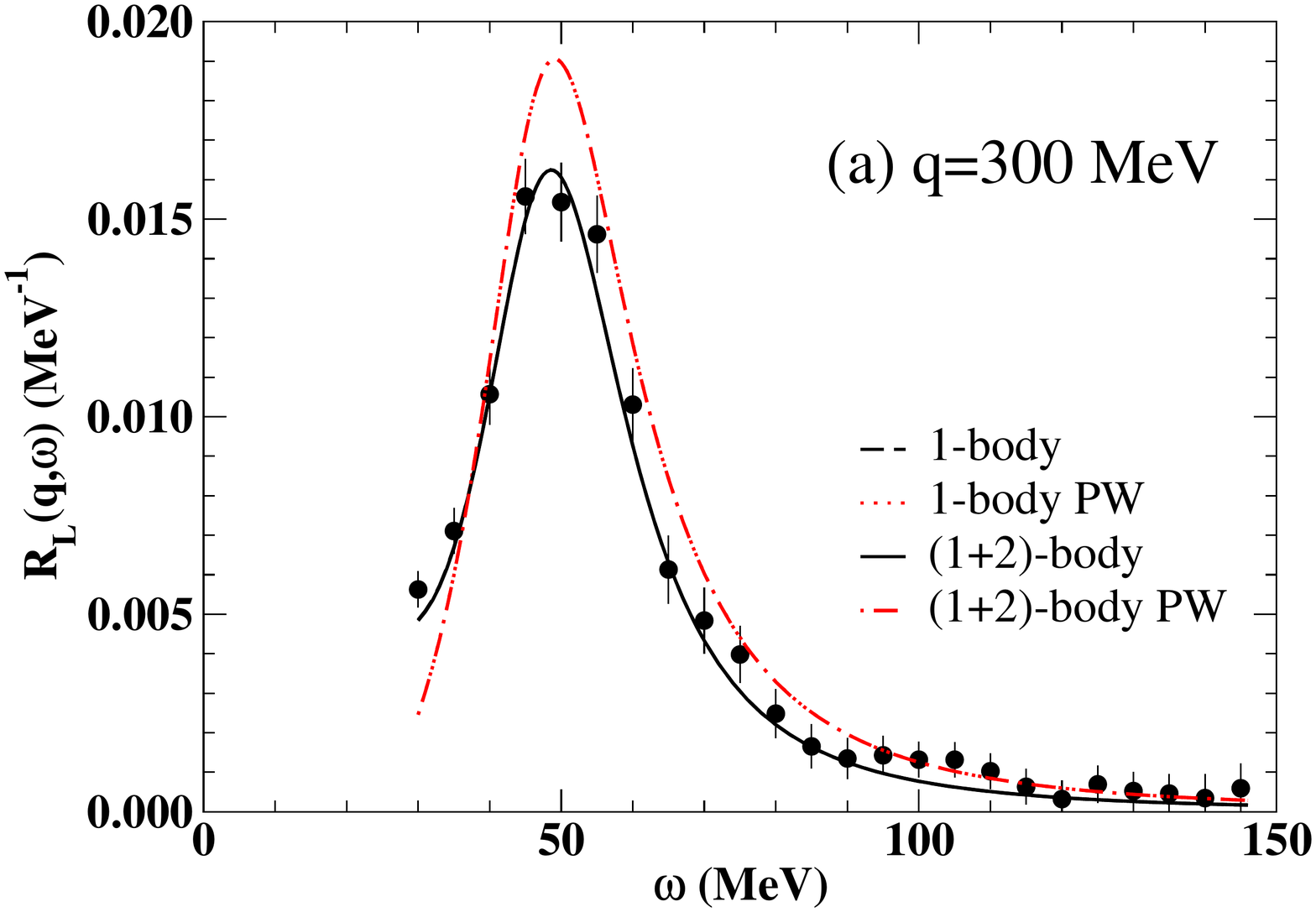}
\includegraphics[width=3in]{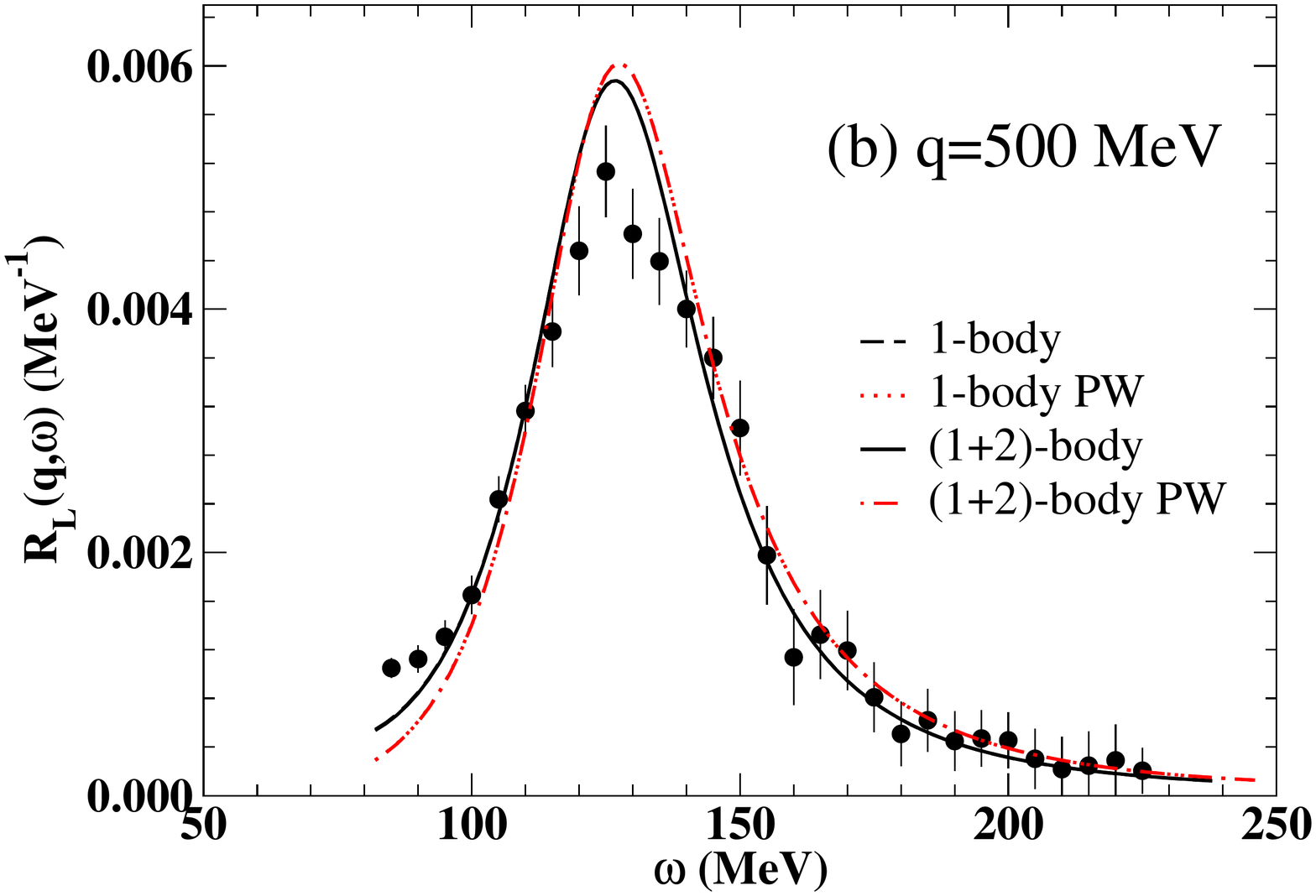}
\includegraphics[width=3in]{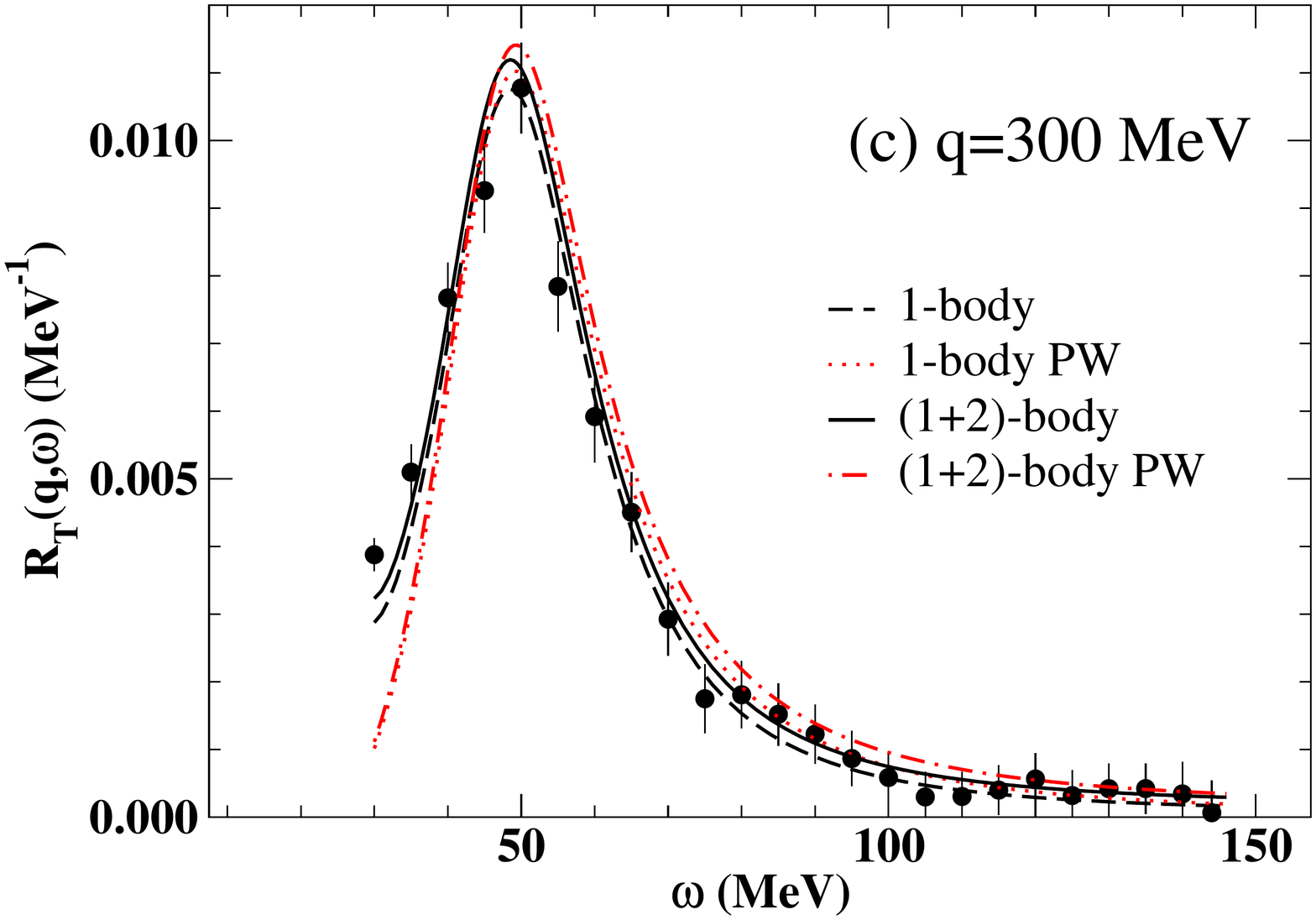}
\includegraphics[width=3in]{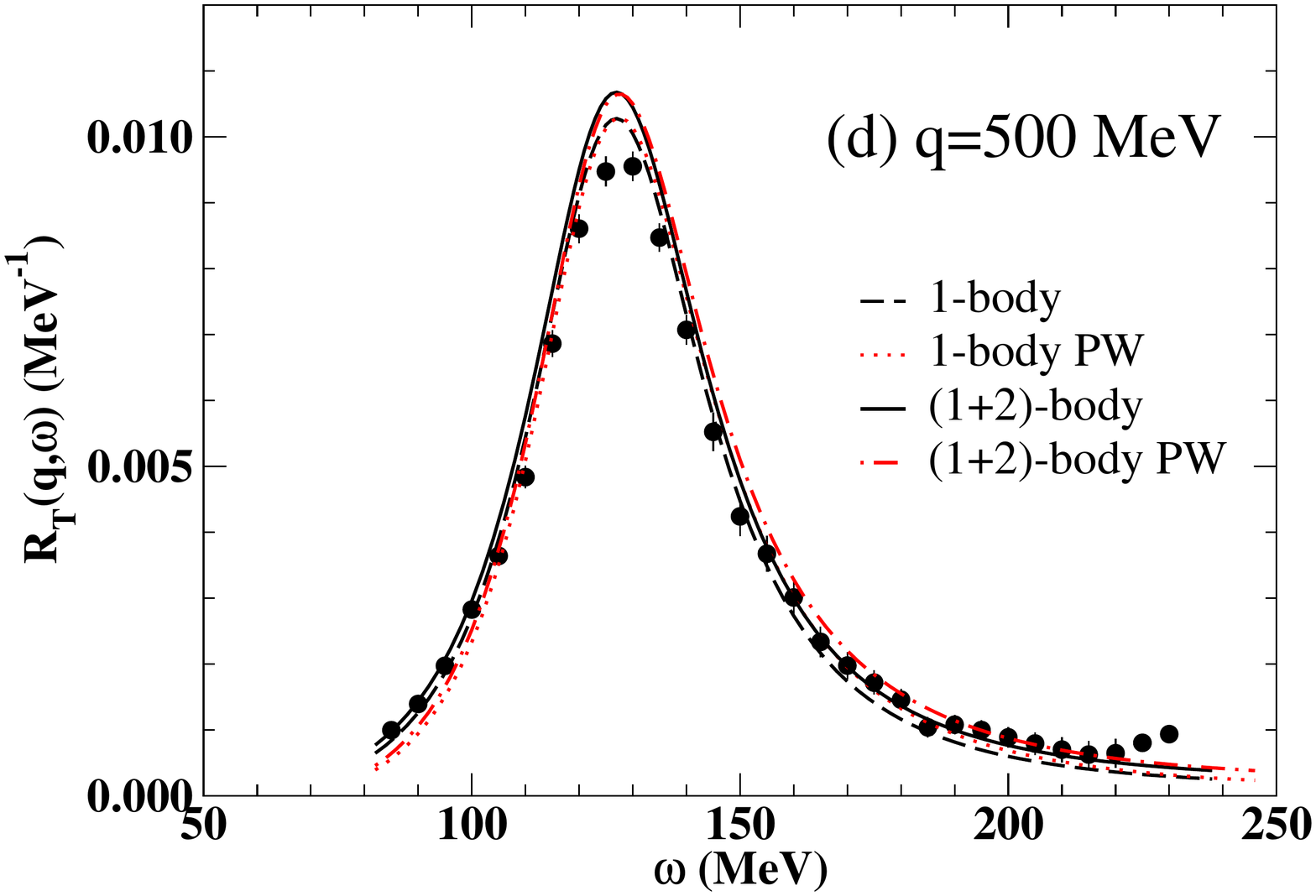}
\caption{(color online) The electromagnetic responses:  longitudinal at $q=300$ MeV (a);  longitudinal
at $q=500$ MeV (b):  transverse at $q=300$ MeV (c);  transverse at $q=500$ MeV (d).  These are
obtained with the AV18 potential
and the inclusion of one-body (dashed line) and (one+two)-body (solid line) terms in the
nuclear electromagnetic charge operator, and are compared to data.  Also shown are the results
obtained with plane-wave (PW) final states.}
\label{fig:em}
\end{figure}

As an additional caveat, we note that the present theory cannot describe the inclusive cross section
in the pion-production region, for example the $\Delta$-excitation peak region, since
no mechanisms for (real) single- or multi-pion production are included in it.  However,
it does reproduce quite well the observed $d(e,e^\prime)$ inclusive cross section
in the quasi-elastic peak region at intermediate values of the three-momentum transfer.
This is illustrated in Fig.~\ref{fig:em} (a-d), where the longitudinal and transverse
response functions $R_L$ and $R_T$ obtained at Bates~\cite{Dytman90} by Rosenbluth
separation of $(e,e^\prime)$ data at momentum transfers of 300 MeV and 500 MeV are
compared with theory.  In these figures, we show separately the response functions calculated
with an electromagnetic current including one-body only and (one+two)-body terms,
as well as those obtained by replacing the fully interacting $NN$ states of Eq.~(\ref{eq:psipw})
with plane waves (curves labeled by PW).  Two-body terms in $R_L$ give negligible contributions,
those in $R_T$ lead to an increase of the transverse strength over the whole quasi-elastic
region, which amounts to a few \% at the top of the peak, but becomes sizable (relative
to the one-body response) as the energy transfer $\omega$ increases well beyond
the quasi-elastic peak energy $\omega_{\rm qe} = (q^2+m^2)^{1/2}-m$.  At these momentum transfers,
the quasi-elastic and $\Delta$ peaks in $R_T$, the latter at $\omega_{\Delta}=(q^2+m_\Delta^2)^{1/2}-m$
($m_\Delta=1232$ MeV), are well separated---note, however, the rise seen in the data
at $q=500$ MeV and the highest $\omega$'s, presumably due to (transverse) strength creeping
in from the $\Delta$-peak region.  Interaction effects in the $NN$ continuum states are
important, particularly at low momentum transfers and/or for energy transfers close to the threshold
for deuteron breakup.  However, at the larger $q$-values plane-wave states provide
response functions in the quasi-elastic region, which are fairly close to those predicted by the
exact scattering states.  Finally, we note that at $q=500$ MeV and quasi-elastic
energies theory over-predicts the measured longitudinal response.  As a consequence,
the total integrated longitudinal strength---the Coulomb sum rule---obtained from these
data~\cite{Schiavilla89} is smaller than calculated.  On the other hand, there is excellent
agreement between the theoretical and measured Coulomb sum rules at $q=300$ (and 400) MeV~\cite{Schiavilla89}.

In Table~\ref{tab:gnu} the columns labeled $\nu_l$-NC and $\overline{\nu}_l$-NC refer to the
NC-induced reactions in Eq.~(\ref{eq:nnr}), those labeled $\nu_e$-CC and $\overline{\nu}_e$-CC
refer to CC-induced reactions in Eq.~(\ref{eq:ncr}), and the initial neutrino energy is between 5 MeV (close to threshold)
and 100 MeV.  In this energy range the cross sections change rapidly, by 3--4 orders of magnitude,
and interaction effects in the final scattering states are important.   Two-body terms in the vector
and axial pieces of the weak current increase the one-body cross section typically by a 2--3 \%
for both the NC- and CC-induced reactions, in agreement with the results of Ref.~\cite{Nakamura02}.

There are differences between the present calculations and those of Nakamura {\it et al.}~\cite{Nakamura02}---mostly
having to do with the model for the weak current---which, however, only lead to small numerical differences
in the predicted cross section values, as shown below.  As we have already remarked in Sec.~\ref{sec:2b}, the authors of
that work ignore the relativistic corrections proportional $1/m^2$ in the one-body axial current~(\ref{eq:j5}),
and use a different short-range parametrization for the two-body vector and axial currents than adopted here.
In addition, the cutoff masses entering the nucleon form factors have slightly different values from those
listed in Sec.~\ref{sec:1b}.  In order to have a more meaningful comparison with the results of that
work, we have carried out a calculation of the NC- and CC-induced cross sections at three representative
initial neutrino energies, in which we have removed the relativistic correction in the one-body axial current and
have changed the cutoff mass values in the nucleon form factors so as to match those used in Ref.~\cite{Nakamura02}.
Inspection of Table~\ref{tab:cmp} shows that the two calculations are typically within less than 1\%
of each other.  This level of agreement should be viewed as satisfactory, given the 
different ways in which the two calculations are carried out in practice.  The authors of
Ref.~\cite{Nakamura02} rely on a multipole expansion of the cross section, whereas
we compute the matrix elements entering the various response functions by direct
numerical integrations, which avoid the need for introducing (cumbersome) multipole expansions
of the weak transition operators.  Of course, the present calculations are computationally
intensive: evaluation of the NC cross sections requires about 40 mins per neutrino
energy on 512 processors, and similar times for each of the two CC cross sections.
\begin{table}[bth]
\caption{Total cross sections in cm$^2$ for the NC-induced processes on the deuteron
as function of the initial neutrino energy $\epsilon$, obtained with the AV18
potential and the inclusion of one-body (1) and (one+two)-body (1+2) terms in the weak current.
Results corresponding to continuum final states (C) and plane wave final states (PW) are listed.} 
\begin{tabular}{c||c|c|c|c||c|c|c|c||}
$\epsilon$ (MeV) & $\nu_l$(1,C) & $\nu_l$(1+2,C) & $\nu_l$(1,PW) & $\nu_l$(1+2,PW) & $\overline{\nu}_l$(1,C) & $\overline{\nu}_l$(1+2,C)
& $\overline{\nu}_l$(1,PW) & $\overline{\nu}_l$(1+2,PW)\\
\hline
         100 &  2.577(--40) &  2.657(--40) &  2.469(--40) &  2.510(--40) &  1.604(--40) &  1.640(--40) &  1.607(--40) &  1.619(--40) \\
         150 &  5.720(--40) &  5.935(--40) &  5.626(--40) &  5.752(--40) &  3.003(--40) &  3.075(--40) &  3.096(--40) &  3.124(--40) \\
         200 &  9.435(--40) &  9.846(--40) &  9.384(--40) &  9.650(--40) &  4.345(--40) &  4.460(--40) &  4.526(--40) &  4.576(--40) \\
         250 &  1.324(--39) &  1.389(--39) &  1.324(--39) &  1.369(--39) &  5.531(--40) &  5.695(--40) &  5.778(--40) &  5.858(--40) \\
         300 &  1.683(--39) &  1.772(--39) &  1.689(--39) &  1.754(--39) &  6.546(--40) &  6.762(--40) &  6.842(--40) &  6.962(--40) \\
         350 &  2.003(--39) &  2.116(--39) &  2.014(--39) &  2.101(--39) &  7.420(--40) &  7.687(--40) &  7.752(--40) &  7.917(--40) \\
         400 &  2.279(--39) &  2.414(--39) &  2.295(--39) &  2.403(--39) &  8.186(--40) &  8.504(--40) &  8.545(--40) &  8.760(--40) \\
450 &  2.509(--39) &  2.664(--39) &  2.531(--39) &  2.660(--39) &  8.856(--40) &  9.221(--40) &  9.255(--40) &  9.520(--40) \\
	     500 &  2.703(--39) &  2.874(--39) &  2.727(--39) &  2.874(--39) &  9.503(--40) &  9.916(--40) &  9.906(--40) &  1.023(--40) \\
         550 &  2.861(--39) &  3.046(--39) &  2.888(--39) &  3.051(--39) &  1.010(--39) &  1.056(--39) &  1.052(--39) &  1.089(--39) \\
         600 &  2.989(--39) &  3.185(--39) &  3.019(--39) &  3.196(--39) &  1.068(--39) &  1.118(--39) &  1.110(--39) &  1.153(--39) \\
         650 &  3.093(--39) &  3.299(--39) &  3.125(--39) &  3.315(--39) &  1.124(--39) &  1.178(--39) &  1.166(--39) &  1.214(--39) \\
         700 &  3.176(--39) &  3.390(--39) &  3.210(--39) &  3.411(--39) &  1.178(--39) &  1.237(--39) &  1.221(--39) &  1.275(--39) \\
         750 &  3.243(--39) &  3.463(--39) &  3.278(--39) &  3.489(--39) &  1.232(--39) &  1.295(--39) &  1.275(--39) &  1.333(--39) \\
         800 &  3.297(--39) &  3.522(--39) &  3.333(--39) &  3.552(--39) &  1.284(--39) &  1.352(--39) &  1.327(--39) &  1.391(--39) \\
         850 &  3.340(--39) &  3.570(--39) &  3.377(--39) &  3.603(--39) &  1.337(--39) &  1.408(--39) &  1.379(--39) &  1.448(--39) \\
         900 &  3.374(--39) &  3.608(--39) &  3.412(--39) &  3.644(--39) &  1.388(--39) &  1.463(--39) &  1.430(--39) &  1.504(--39) \\
         950 &  3.403(--39) &  3.639(--39) &  3.440(--39) &  3.678(--39) &  1.440(--39) &  1.518(--39) &  1.481(--39) &  1.559(--39) \\
        1000 &  3.425(--39) &  3.663(--39) &  3.461(--39) &  3.704(--39) &  1.490(--39) &  1.572(--39) &  1.530(--39) &  1.613(--39) \\
\hline
\end{tabular}
\label{tab:nc}
\end{table}
\begin{table}[bth]
\caption{Same as in Table~\protect\ref{tab:nc}, but for CC-induced processes.} 
\begin{tabular}{c||c|c|c|c||c|c|c|c||}
$\epsilon$ (MeV) & $\overline{\nu}_e$(1,C) & $\overline{\nu}_e$(1+2,C)
& $\overline{\nu}_e$(1,PW) & $\overline{\nu}_e$(1+2,PW)& $\nu_e$(1,C) & $\nu_e$(1+2,C) & $\nu_e$(1,PW) & $\nu_e$(1+2,PW) \\
\hline
         100 &  2.567(--40) &  2.606(--40) &  2.362(--40) &  2.370(--40) &  6.424(--40) &  6.631(--40) &  5.908(--40) &  6.023(--40) \\
         150 &  4.688(--40) &  4.751(--40) &  4.487(--40) &  4.491(--40) &  1.516(--39) &  1.574(--39) &  1.440(--39) &  1.477(--39) \\
         200 &  6.736(--40) &  6.830(--40) &  6.555(--40) &  6.568(--40) &  2.605(--39) &  2.719(--39) &  2.525(--39) &  2.603(--39) \\
        250 &  8.677(--40) &  8.822(--40) &  8.520(--40) &  8.567(--40) &  3.775(--39) &  3.958(--39) &  3.699(--39) &  3.833(--39) \\
        300 &  1.059(--39) &  1.082(--39) &  1.044(--39) &  1.056(--39) &  4.928(--39) &  5.186(--39) &  4.854(--39) &  5.052(--39) \\
         350 &  1.254(--39) &  1.286(--39) &  1.239(--39) &  1.261(--39) &  5.981(--39) &  6.315(--39) &  5.923(--39) &  6.189(--39) \\
         400 &  1.455(--39) &  1.499(--39) &  1.441(--39) &  1.475(--39) &  6.920(--39) &  7.320(--39) &  6.876(--39) &  7.210(--39) \\
         450 &  1.663(--39) &  1.722(--39) &  1.650(--39) &  1.698(--39) &  7.778(--39) &  8.248(--39) &  7.704(--39) &  8.102(--39) \\
         500 &  1.879(--39) &  1.952(--39) &  1.865(--39) &  1.930(--39) &  8.524(--39) &  9.053(--39) &  8.410(--39) &  8.868(--39) \\
         550 &  2.100(--39) &  2.189(--39) &  2.087(--39) &  2.169(--39) &  9.064(--39) &  9.636(--39) &  9.005(--39) &  9.519(--39)   \\
         600 &  2.323(--39) &  2.428(--39) &  2.309(--39) &  2.410(--39) &  9.556(--39) &  1.017(--38) &  9.504(--39) &  1.007(--38)  \\
         650 &  2.548(--39) &  2.671(--39) &  2.537(--39) &  2.656(--39) &  9.966(--39) &  1.062(--38) &  9.920(--39) &  1.053(--38)   \\
         700 &  2.777(--39) &  2.916(--39) &  2.766(--39) &  2.905(--39) &  1.031(--38) &  1.098(--38) &  1.027(--38) &  1.091(--38)   \\
         750 &  3.005(--39) &  3.161(--39) &  2.995(--39) &  3.152(--39) &  1.059(--38) &  1.129(--38) &  1.055(--38) &  1.124(--38)   \\
         800 &  3.232(--39) &  3.403(--39) &  3.223(--39) &  3.399(--39) &  1.082(--38) &  1.154(--38) &  1.079(--38) &  1.150(--38)   \\
         850 &  3.456(--39) &  3.645(--39) &  3.448(--39) &  3.643(--39) &  1.101(--38) &  1.176(--38) &  1.099(--38) &  1.173(--38)   \\
         900 &  3.678(--39) &  3.882(--39) &  3.671(--39) &  3.885(--39) &  1.118(--38) &  1.193(--38) &  1.116(--38) &  1.192(--38)   \\
         950 &  3.896(--39) &  4.116(--39) &  3.890(--39) &  4.122(--39) &  1.131(--38) &  1.208(--38) &  1.129(--38) &  1.208(--38)   \\
        1000 &  4.109(--39) &  4.343(--39) &  4.105(--39) &  4.356(--39) &  1.142(--38) &  1.221(--38) &  1.141(--38) &  1.222(--38)   \\ 
\hline
\end{tabular}
\label{tab:cc}
\end{table}
\begin{figure}[bth]
\includegraphics[width=6in]{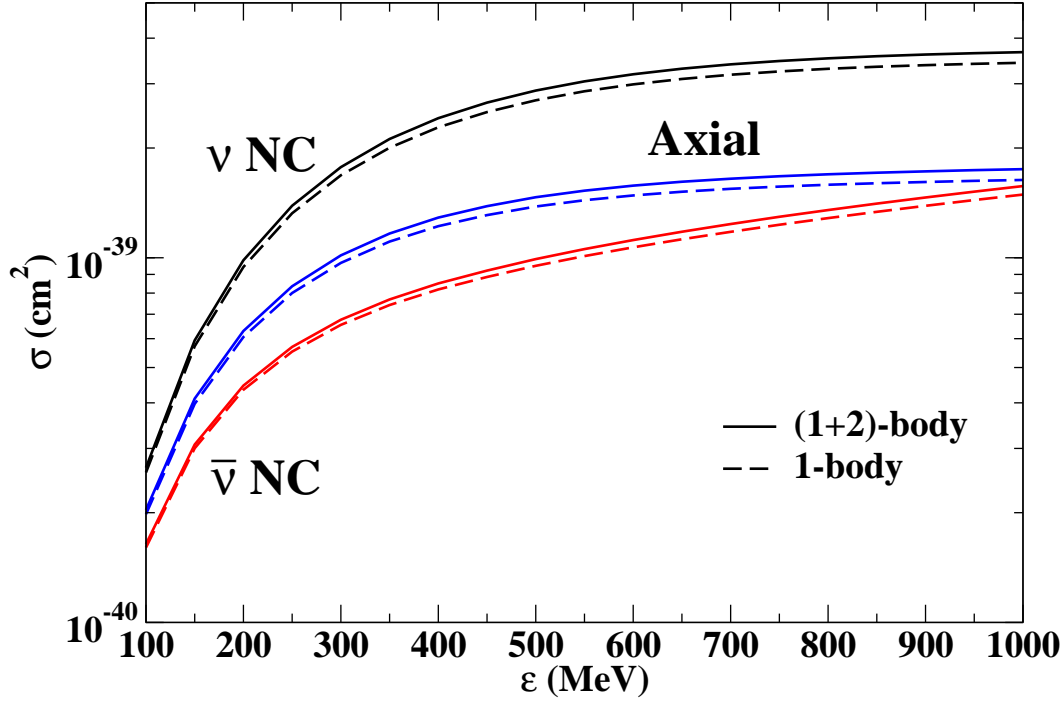}\\
\caption{(color online) Total cross sections for NC-induced processes on the deuteron, obtained with the AV18 potential
and the inclusion of one-body (dashed line) and (one+two)-body (solid line) terms in the weak current.  Also shown
are the total cross sections obtained by retaining only the axial piece of the weak current. See text for explanation.}
\label{fig:nc1+2}
\end{figure}
\begin{figure}[bth]
\includegraphics[width=6in]{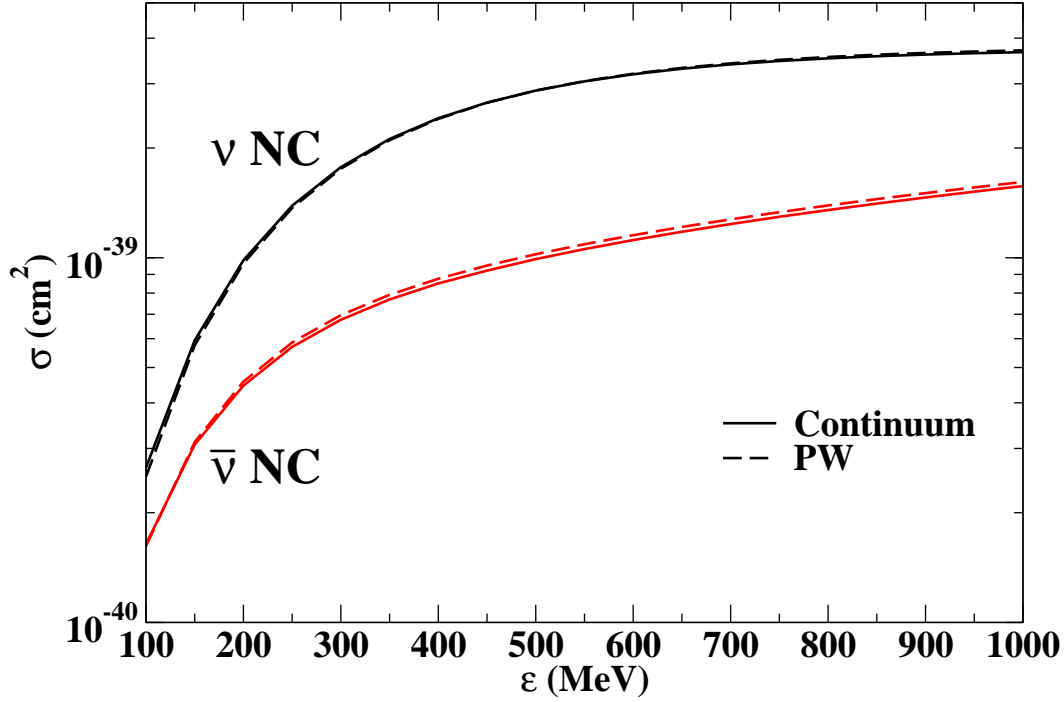}\\
\caption{(color online) Total cross sections for NC-induced processes on the deuteron, obtained with the AV18 potential
and the inclusion of (one+two)-body terms in the weak current.  Also shown are the total cross sections
obtained with plane-wave (PW) final states.}
\label{fig:ncx}
\end{figure}
\begin{figure}[bth]
\includegraphics[width=6in]{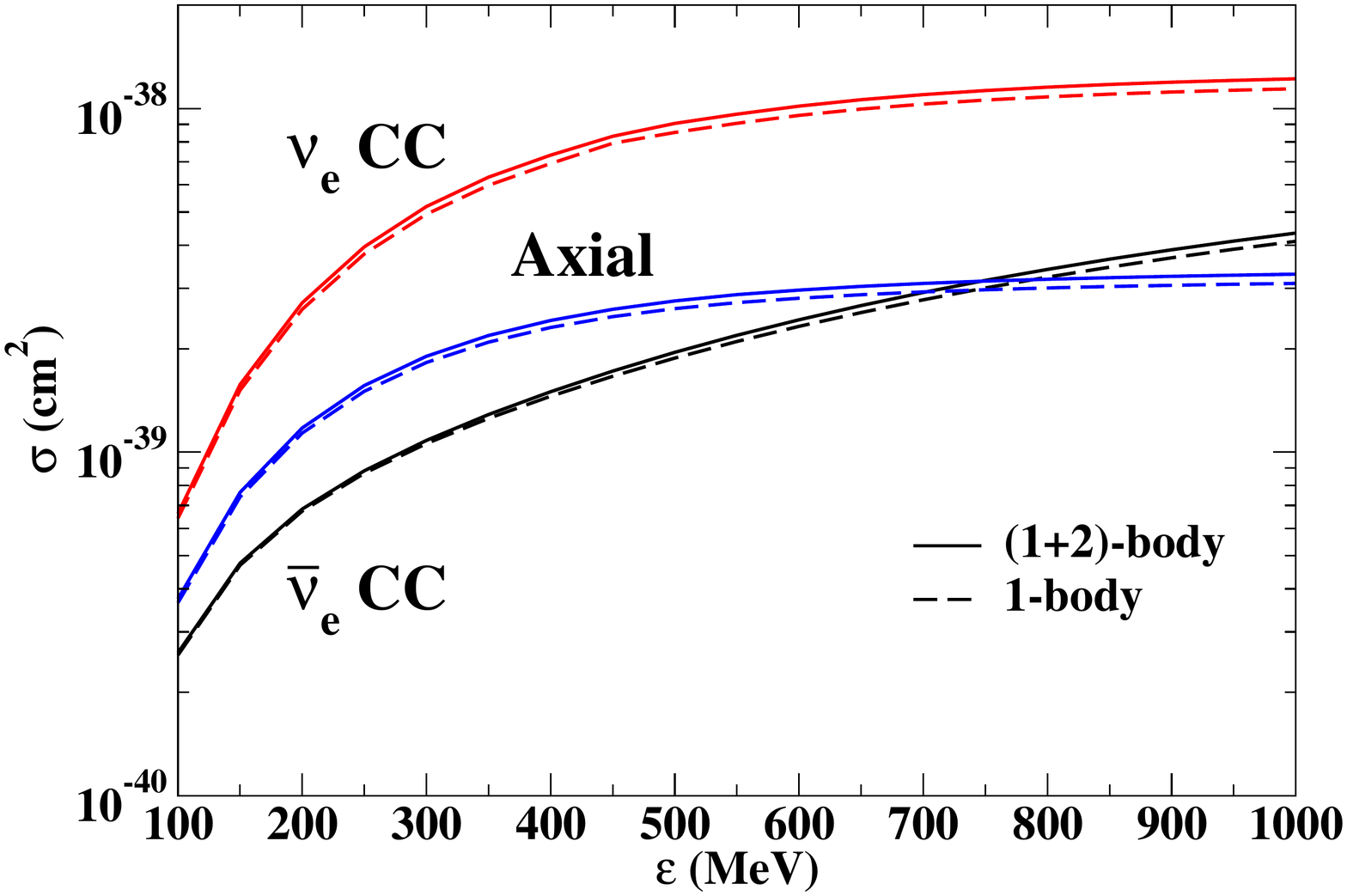}\\
\caption{(color online) Same as in Fig.~\protect\ref{fig:nc1+2}, but for CC-induced processes on the deuteron.}
\label{fig:cc1+2}
\end{figure}
\begin{figure}[bth]
\includegraphics[width=6in]{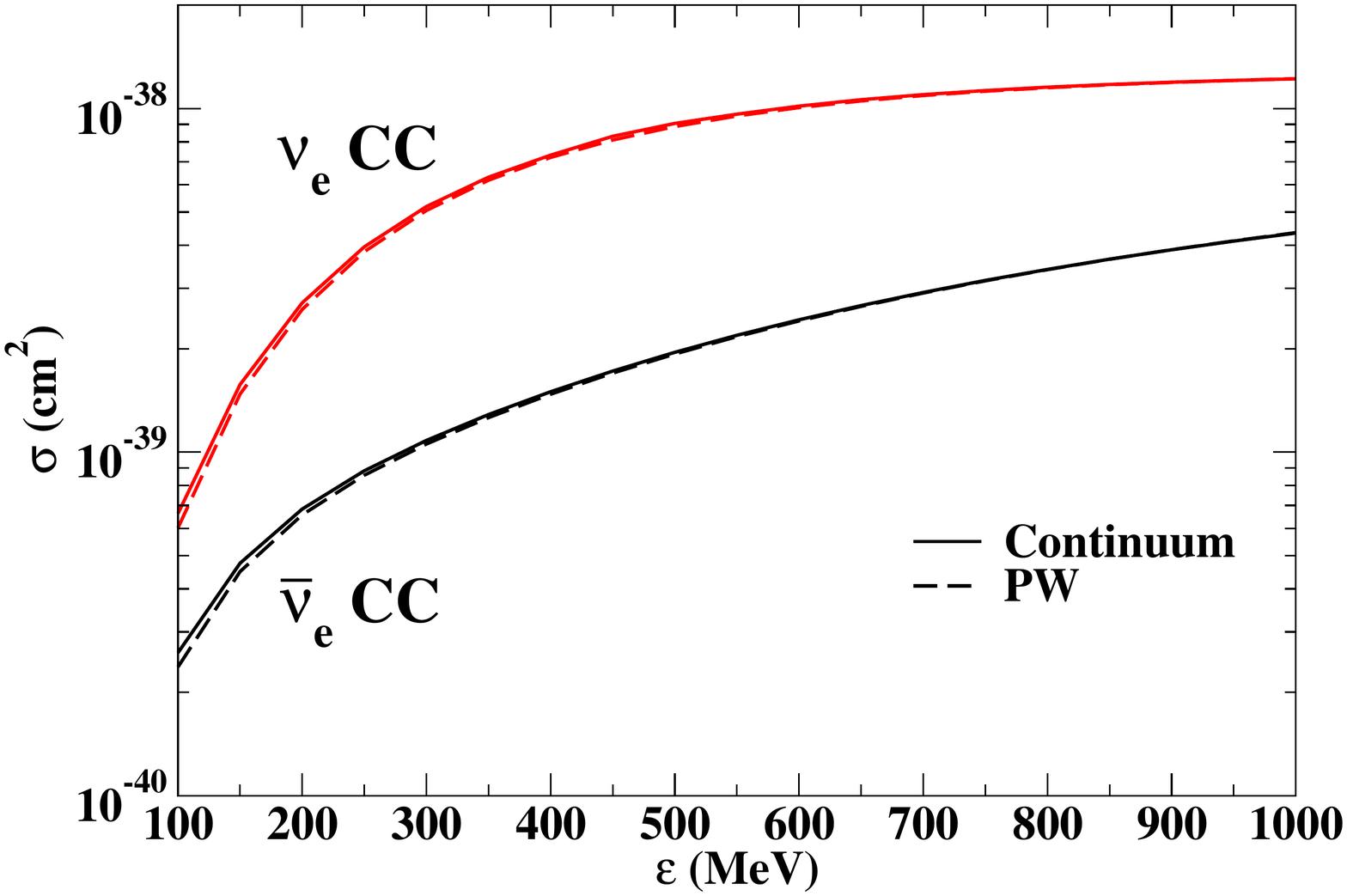}\\
\caption{(color online) Same as in Fig.~\protect\ref{fig:ncx}, but for CC-induced processes on the deuteron.}
\label{fig:ccx}
\end{figure}
\begin{table}[bth]
\caption{(color on line) Total cross sections in cm$^2$ for the NC- and CC-induced processes on the
deuteron obtained in Ref.~\protect\cite{Nakamura02} and in the present work at selected
initial neutrino energies.  Note that the values under the heading
``this work'' are slightly different than those reported in Table~\protect\ref{tab:gnu}
for the reasons explained in the text.}
\begin{tabular}{c|c|c|c|c|c|c|}
 $\epsilon$ (MeV)& \multicolumn{2}{c}  {5}  &  \multicolumn{2}{c} {50} & \multicolumn{2}{c}  {100} \\
 \hline\hline
  & Ref.~\protect\cite{Nakamura02}& this work & Ref.~\protect\cite{Nakamura02}& this work &
 Ref.~\protect\cite{Nakamura02}& this work \\
\hline
$\nu_l$-NC &9.570(--44) &9.601(--44) & 5.944(--41)& 5.942(--41)& 2.711(--40)& 2.703(--40)\\
$\overline{\nu}_l$-NC &9.364(--44) &9.403(--44) &4.535(--41) &4.589(--41) &1.647(--40) & 1.674(--40)\\
$\nu_e$-CC & 3.463(--43)&3.440(--43) & 1.376(--40) & 1.367(--40)& 6.836(--40)&6.735(--40) \\
$\overline{\nu}_e$-CC &2.836(--44) &2.842(--44) & 7.372(--41) & 7.475(--41)&2.618(--40) &2.659(--40) \\
\hline\hline
\end{tabular}
\label{tab:cmp}
\end{table}

The calculated cross sections for the NC- and CC-induced reactions are listed, respectively, in Tables~\ref{tab:nc}
and~\ref{tab:cc} for incident neutrino energies between 100 MeV and 1000 MeV.  The columns labeled
(1,C) and (1+2,C) [(1,PW) and (1+2,PW)] show results obtained by including fully interacting [plane-wave]
$NN$ final states, and one-body only or (one+two)-body terms in the weak current.  These results are also
displayed in Figs.~\ref{fig:nc1+2}--\ref{fig:ccx}.  The two-body contributions are small, less than
10\% over the whole energy range.  Interaction effects in the final states are found to be even smaller,
which suggests that realistic estimates for these cross sections on the deuteron (and possibly
nuclei with $A>2$) at high energies may be obtained by approximating the final nuclear states by
plane waves, i.e.~by employing the nucleon momentum distribution in the deuteron (or the spectral
function in $A>2$ nuclei).  Lastly, in Figs.~\ref{fig:nc1+2} and~\ref{fig:cc1+2} we also
show the results obtained by including only the axial piece in the weak current.  In this case,
the interference response function $R_{xy}$ vanishes, and consequently the $\nu_l$ and $\overline{\nu}_l$
cross sections are the same. For CC-induced reactions, due to the charge-dependence of the $NN$ final state
($pp$ or $nn$) the neutrino-induced CC reactions has slightly larger cross section (a few \%) even with only
the axial piece in the weak current.  We display the axial contribution in the anti-neutrino-induced CC reactions
in Fig. \ref{fig:cc1+2}.  Axial contributions are larger than vector at low energy $\lesssim (400$--500) MeV, but
become smaller than vector at higher energy.

\begin{table}[bth]
\caption{Total cross sections in cm$^2$ for the NC- and CC-induced processes on the deuteron
at selected initial neutrino energies, obtained with the AV18 potential and the inclusion of one-body
and set I or set II two-body terms in the weak current.} 
\begin{tabular}{c|c|c||c|c||c|c||c|c||}
         & \multicolumn{2}{c} {$\nu_l$-NC} & \multicolumn{2}{c} {$\overline{\nu}_l$-NC} &
\multicolumn{2}{c} {$\nu_e$-CC} & \multicolumn{2}{c} {$\overline{\nu}_e$-CC}  \\
\hline\hline
 $\epsilon$ (MeV)    & set I   & set II &  set I & set II & set I & set II & set I & set II \\
    \hline
     5  &  9.561(--44) & 9.541(--44)    & 9.363(--44) & 9.344(--44) & 3.427(--43)  &  3.421(--43) & 2.831(--44) & 2.826(--44)  \\
    50 &  5.892(--41) & 5.873(--41)   & 4.546(--41)  & 4.530(--41) & 1.348(--40)  &  1.353(--40) & 7.403(--41) & 7.380(--41) \\
  100 &  2.657(--40) & 2.652(--40)   & 1.640(--40)  & 1.636(--40) & 6.631(--40)  &  6.621(--40) & 2.606(--40) & 2.600(--40)  \\
 \hline\hline
\end{tabular}
\label{tab:cmec}
\end{table}
The sensitivity of the results on the model used for the
two-body axial current (Set I or Set II) and $NN$ potential
(AV18 or CD-Bonn) is investigated, respectively,
in Tables~\ref{tab:cmec} and~\ref{tab:cnnp}.  In both cases,
the model dependence is found to be negligible.  The two-body vector currents are taken from the
AV18, and therefore their short-range behavior is not consistent with the CD-Bonn interaction.  This
inconsistency, though, is of little numerical import.  Furthermore, since interaction effects in the
two-nucleon continuum appear to be negligible for neutrino energies $\gtrsim 100$ MeV, the agreement
between the calculated cross sections with the AV18 and CD-Bonn merely reflects the fact that
the momentum distributions predicted by these two potential models are very close to each other
for relative momenta $\lesssim $ 400 MeV.
\begin{table}[bth]
\caption{Total cross sections in cm$^2$ for the NC-induced processes on the deuteron
at selected initial neutrino energies, obtained with the AV18 or CDB potentials and
the inclusion of one-body terms (1) only and both one- and two-body terms (1+2) in the weak current.} 
\begin{tabular}{c||c|c|c|c||c|c|c|c||}
          & \multicolumn{4}{c} {$\nu_l$-NC} & \multicolumn{4}{c} {$\overline{\nu}_l$-NC}  \\
\hline\hline
 $\epsilon$ (MeV)  & AV18(1) &  CDB(1)  & AV18(1+2)  & CDB(1+2) & AV18(1) & CDB(1)&  AV18(1+2) & CDB(1+2)  \\
    \hline
       50  & 5.747(--41) & 5.791(--40)    &  5.892(--41) &  5.847(--40)   & 4.449(--41) & 4.484(--40)  & 4.546(--41)   & 4.519(--40)   \\
     100  & 2.577(--40) & 2.597(--40)   &  2.657(--40) & 2.638(--40)   & 1.604(--40) & 1.617(--40)  &  1.640(--40)  &  1.633(--40)  \\
     500  & 2.703(--39) & 2.715(--39)   &  2.874(--39) & 2.858(--39)   & 9.503(--40) & 9.553(--40)  &   9.916(--40)  &  9.895(--40)   \\
    1000 & 3.425(--39) &  3.442(--39)  &  3.663(--39) & 3.659(--39)   & 1.490(--39) & 1.496(--39)  &   1.572(--39)   &  1.572(--39)   \\
 \hline\hline
\end{tabular}
\label{tab:cnnp}
\end{table}

\begin{figure}[bth]
\includegraphics[width=6in]{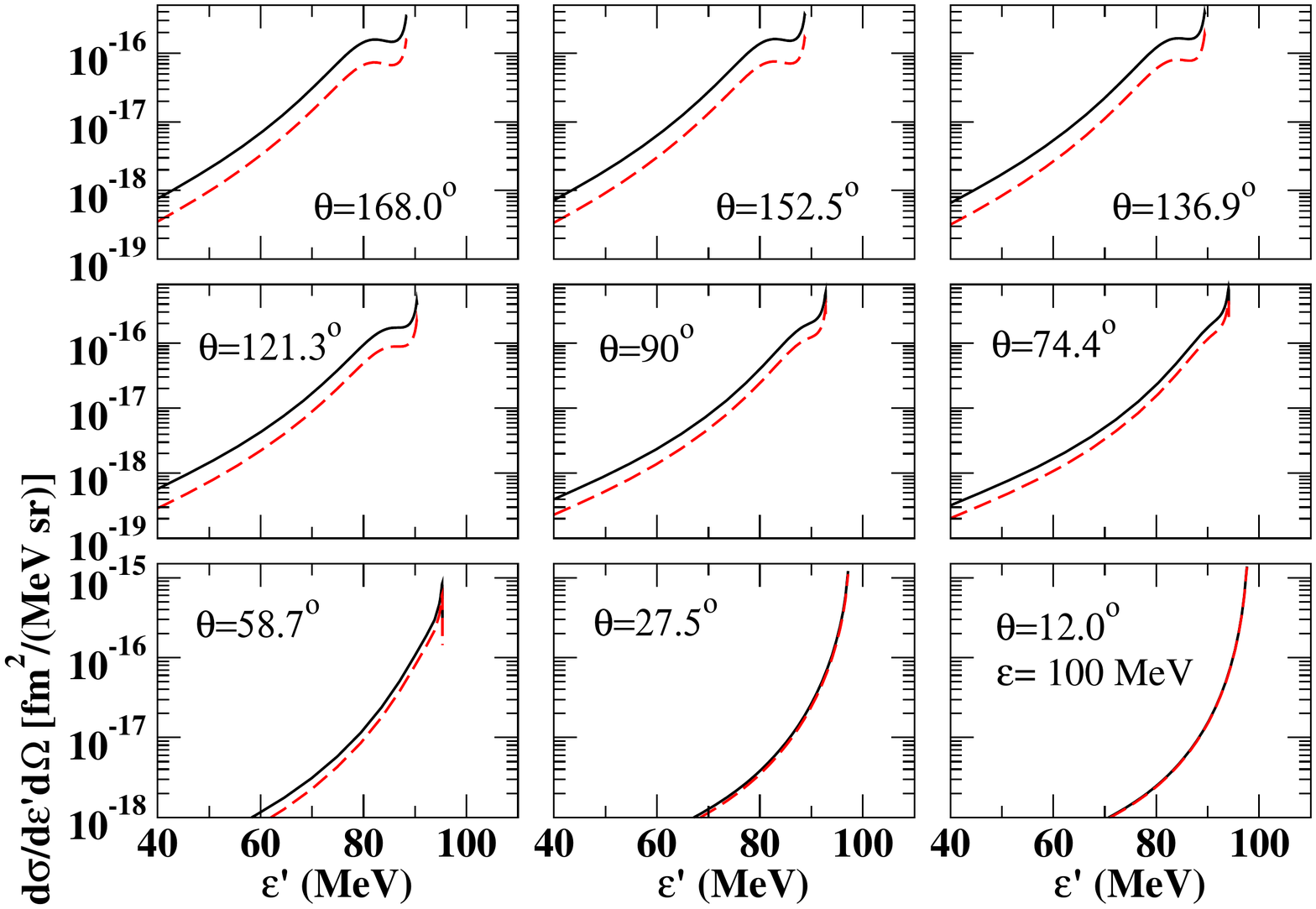}\\
\caption{(color online) Differential cross section for NC-induced processes on the deuteron, obtained with the AV18 potential
and the inclusion of one- and two-body terms in the nuclear weak current, as function of final lepton energy. The incident neutrino energy is 100 MeV. The final lepton angle is indicated in each panel. The (black) solid curve is for neutrino induced processes. The (red) dashed curve is for anti-neutrino induced processes.}
\label{fig:nc100}
\end{figure}

\begin{figure}[bth]
\includegraphics[width=6in]{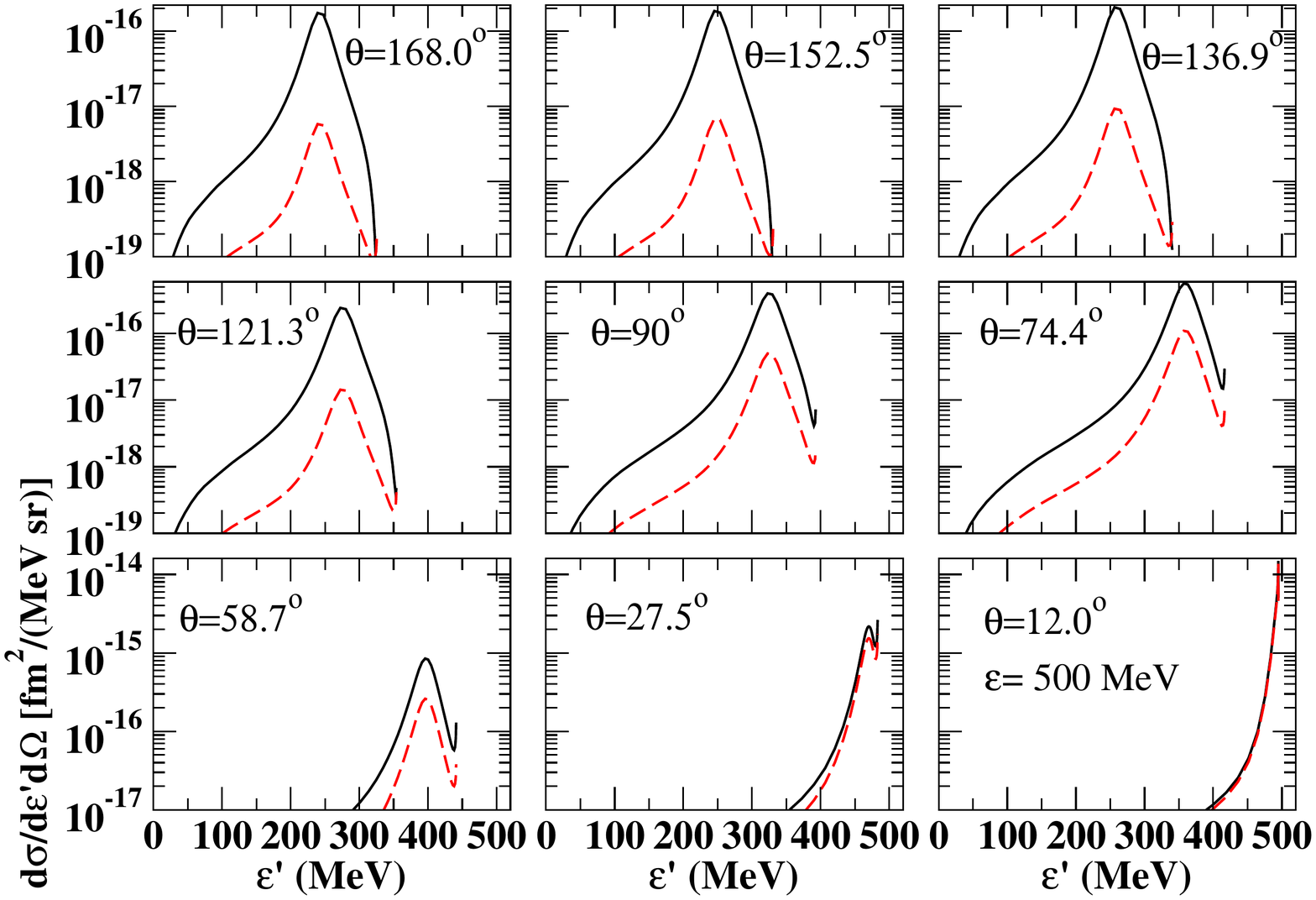}\\
\caption{(color online) Same as Fig. \ref{fig:nc100}, but the incident neutrino energy is 500 MeV.}
\label{fig:nc500}
\end{figure}

\begin{figure}[bth]
\includegraphics[width=6in]{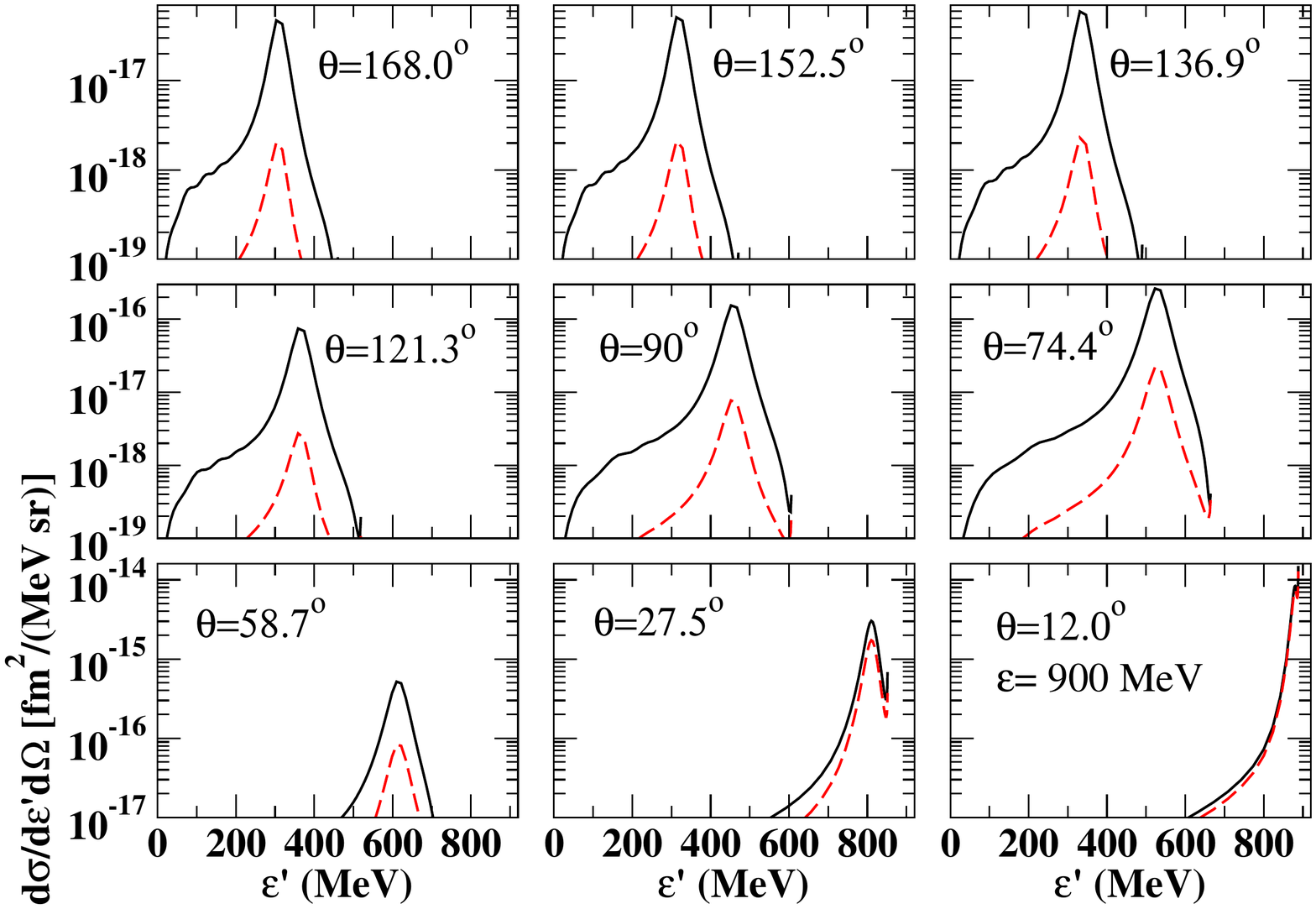}\\
\caption{(color online) Same as Fig. \ref{fig:nc100}, but the incident neutrino energy is 900 MeV.}
\label{fig:nc900}
\end{figure}

\begin{figure}[bth]
\includegraphics[width=6in]{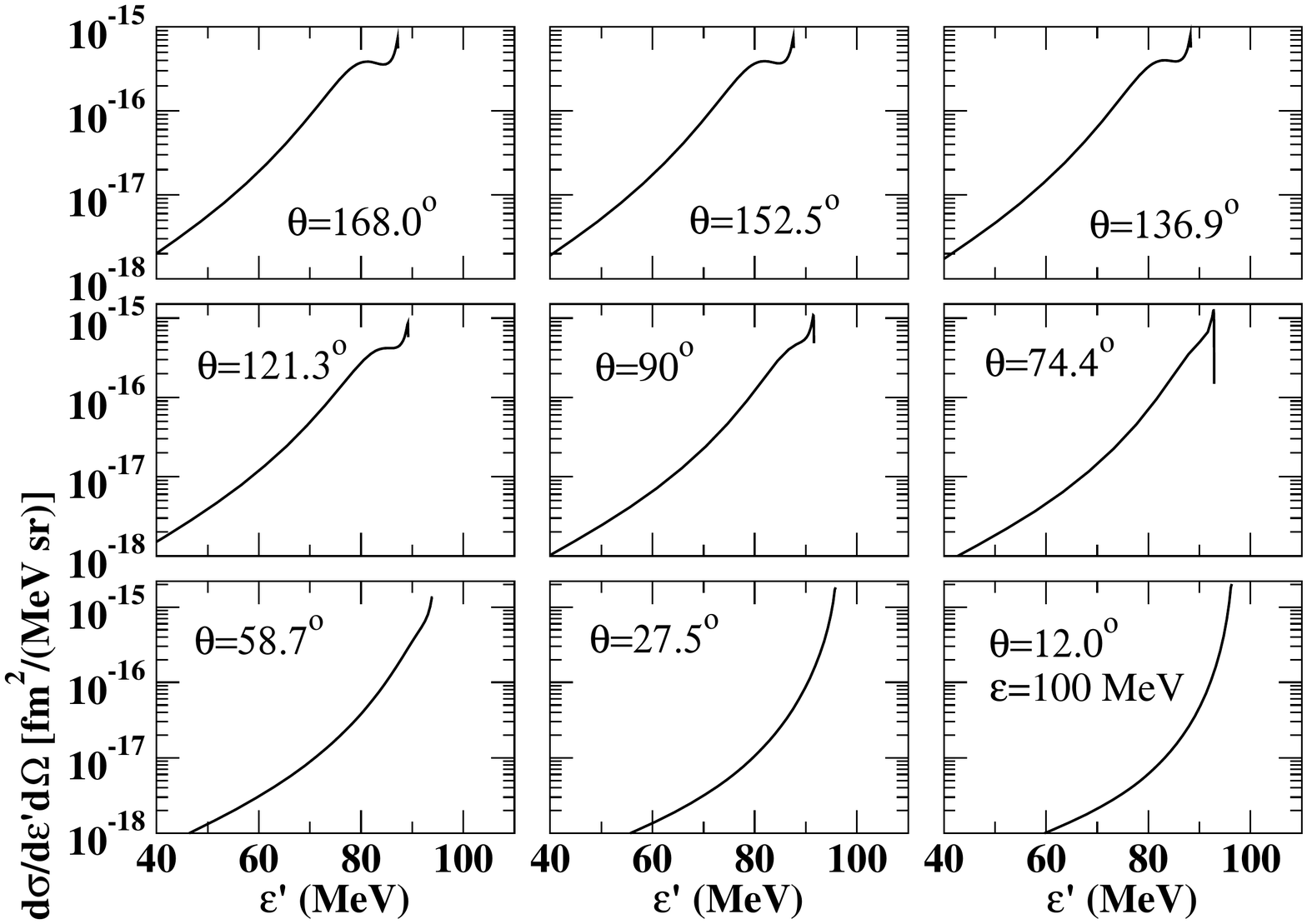}\\
\caption{ Differential cross section for electron anti-neutrino induced CC processes on the deuteron, obtained with the AV18 potential
and the inclusion of one- and two-body terms in the nuclear weak current, as function of final lepton energy. The incident anti-neutrino energy is 100 MeV. The final lepton angle is indicated in each panel.}
\label{fig:nn100}
\end{figure}

\begin{figure}[bth]
\includegraphics[width=6in]{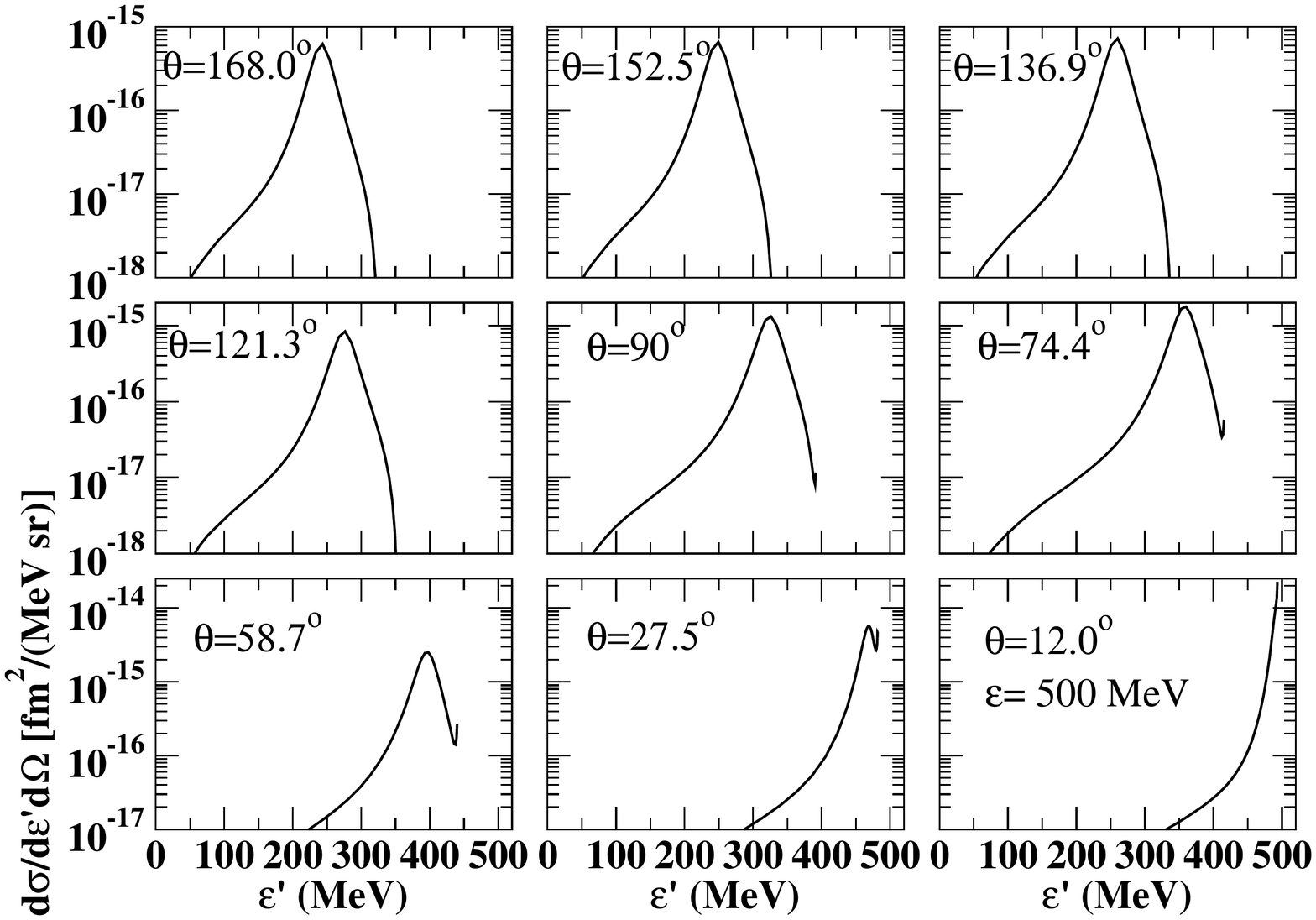}\\
\caption{ Same as Fig. \ref{fig:nn100}, but the incident neutrino energy is 500 MeV.}
\label{fig:nn500}
\end{figure}

\begin{figure}[bth]
\includegraphics[width=6in]{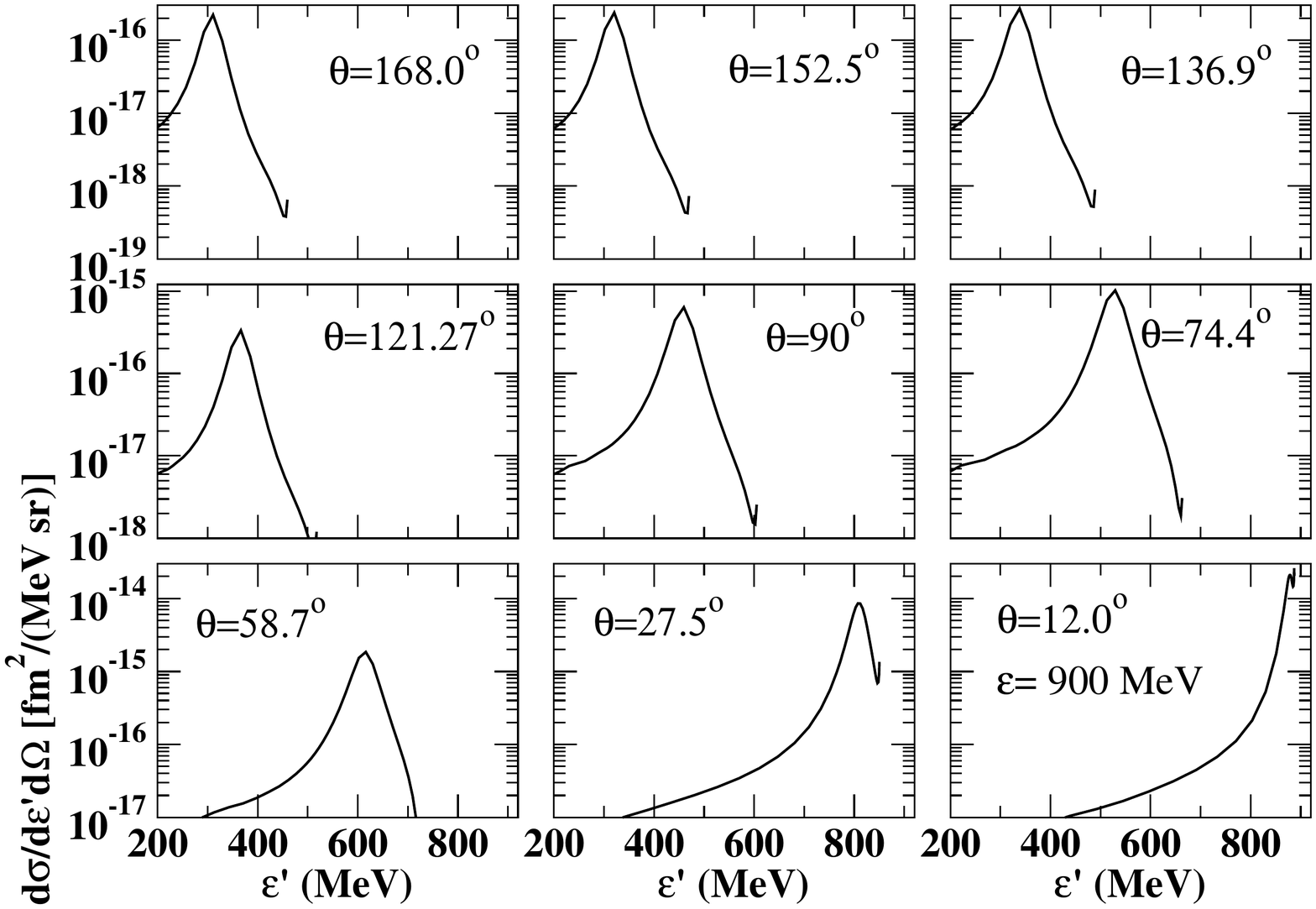}\\
\caption{ Same as Fig. \ref{fig:nn100}, but the incident neutrino energy is 900 MeV.}
\label{fig:nn900}
\end{figure}

\begin{figure}[bth]
\includegraphics[width=6in]{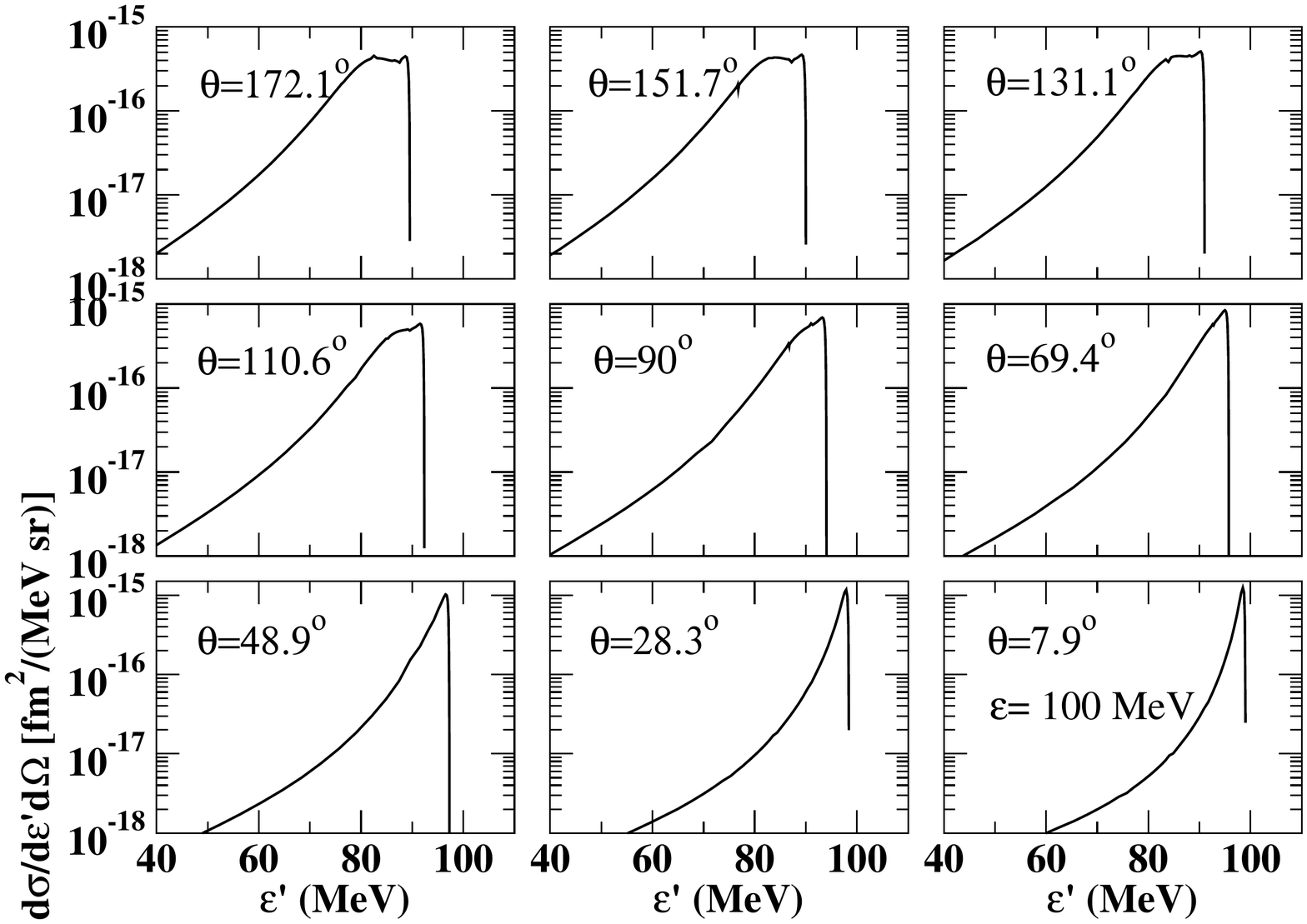}\\
\caption{ Differential cross section for electron neutrino induced CC processes on the deuteron, obtained with the AV18 potential
and the inclusion of one- and two-body terms in the nuclear weak current, as function of final lepton energy. The incident neutrino energy is 100 MeV. The final lepton angle is indicated in each panel.}
\label{fig:pp100}
\end{figure}

\begin{figure}[bth]
\includegraphics[width=6in]{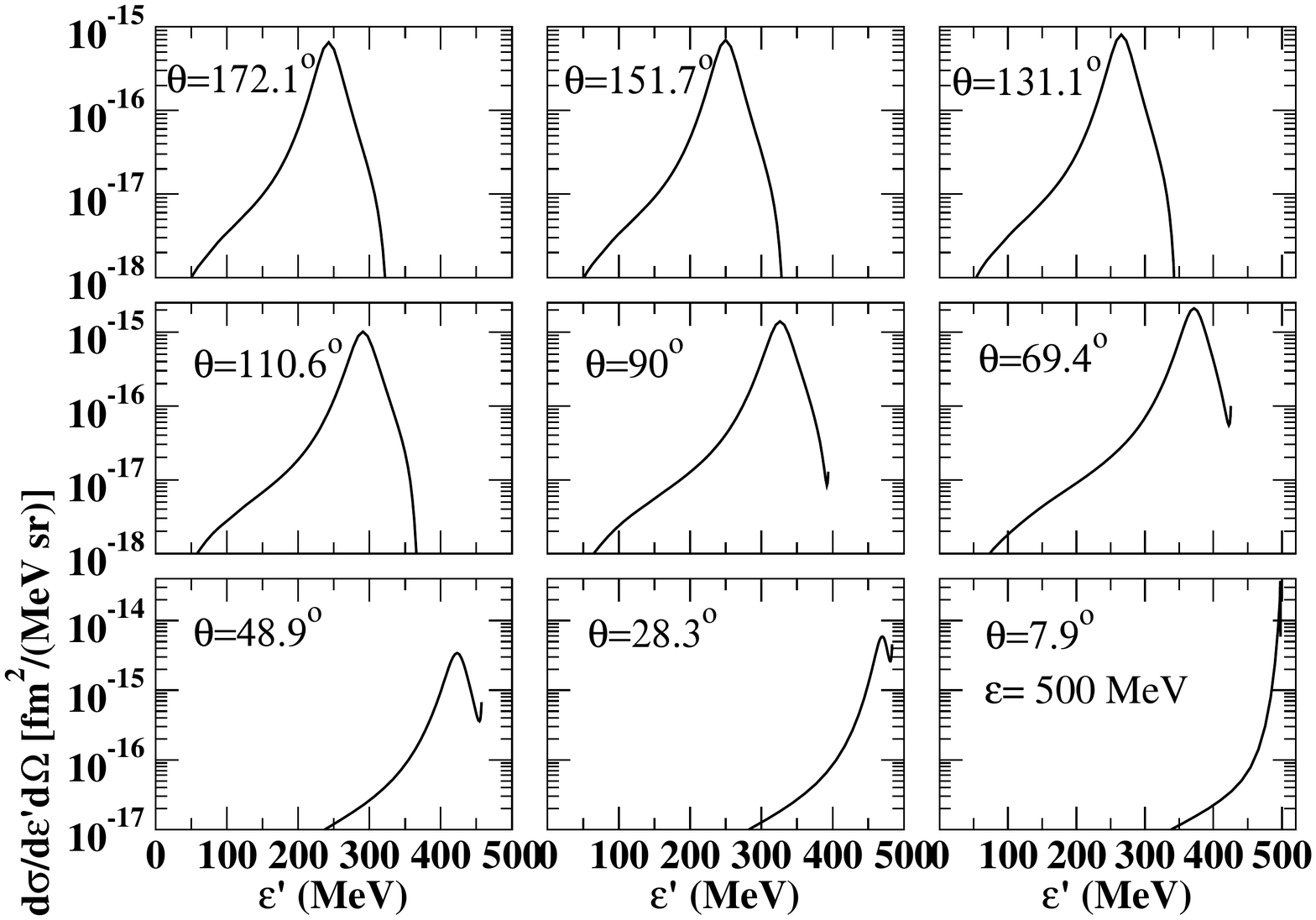}\\
\caption{ Same as Fig. \ref{fig:pp100}, but the incident neutrino energy is 500 MeV.}
\label{fig:pp500}
\end{figure}

\begin{figure}[bth]
\includegraphics[width=6in]{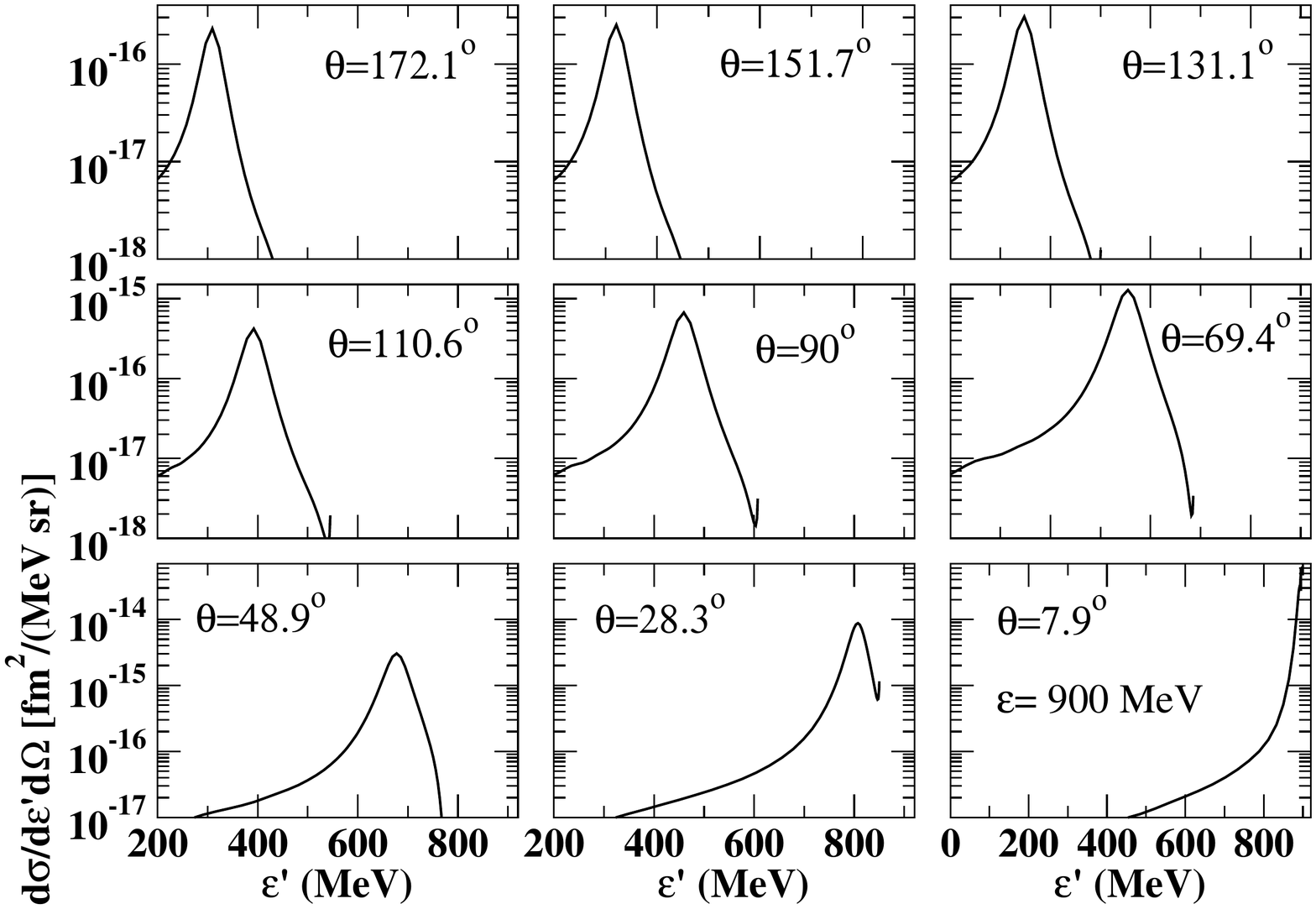}\\
\caption{ Same as Fig. \ref{fig:pp100}, but the incident neutrino energy is 900 MeV.}
\label{fig:pp900}
\end{figure}

In Figs.~\ref{fig:nc100}--\ref{fig:pp900} we show the differential cross sections
for the NC- and CC-induced reactions as function of the final lepton energy $\epsilon^\prime$
and scattering angle $\theta$ at three incident neutrino energies, $\epsilon=100,500, 900$ MeV. 
The quasi-elastic peak is located at a final energy $\epsilon^\prime_{\rm qe}$ given by
\begin{equation}
\epsilon^\prime_{\rm qe} =\frac{\epsilon}{1+(2\, \epsilon/m)\, {\rm sin}^2\theta/2} \ ,
\label{eq:eqe}
\end{equation}
where we have neglected the lepton mass in the case of the CC processes.
Therefore as $\theta$ changes from the backward to the forward hemisphere, the quasi-elastic
peak moves to the right, i.e.~towards higher and higher energies.  Indeed, at forward angles
it merges with the threshold peak due to the quasi-bound $^1$S$_0$ state.  This
latter peak is very pronounced at low $\epsilon$, but becomes more and
more suppressed by the form factor $\sim \langle ^1{\rm S}_0\!\mid\! j_0(q\, r/2)\!\mid\! d\rangle$
as $\epsilon$ increases. 

Finally, it is interesting to compare the results above with those obtained in a naive
model, in which the deuteron is taken to consist of a free proton and neutron initially at rest.
The lab-frame cross sections of the NC-induced processes on the nucleon, and of the CC-induced processes
$n(\nu_e,e^-)p$ and  $p(\overline{\nu}_e,e^+)n$ in the limit in which
the final electron/positron mass and proton-neutron mass difference are neglected, read~\cite{Thomas01}:
\begin{equation}
\left(\frac{ {\rm d}\sigma}{ {\rm d}\epsilon^\prime {\rm d}\Omega}\right)^{\rm NC/CC}_{\nu/\overline{\nu}}
= \frac{G^2\, m^2}{8\pi^2}\, \left( \frac{\epsilon^\prime}{\epsilon}\right)^2 \,
\delta(\epsilon^\prime-\epsilon^\prime_{\rm qe})
\left[ A^{\rm NC/CC}\mp \frac{s-u}{m^2} B^{\rm NC/CC} + \frac{(s-u)^2}{m^4}\, C^{\rm NC/CC} \right] \ ,
\end{equation}
where $G$=$G_F $ or $G_F \, {\rm cos}\, \theta_C$ for NC or CC, the $-$ ($+$) sign in the
second term is relative to the $\nu$ ($\overline{\nu}$) initiated reactions, $\epsilon^\prime_{\rm qe}$
has been defined in Eq.~(\ref{eq:eqe}), and $s-u=4\, m\,\epsilon -Q^2$ with $Q^2=4\, \epsilon\, \epsilon^\prime
\, \sin^2\theta/2$.  The structure functions $A(Q^2)$, $B(Q^2)$, and $C(Q^2)$ for both NC and CC
are given in terms of nucleon form factors in Appendix~\ref{app:app2}.

\begin{figure}[bth]
 \centering
\includegraphics[width=3.5in]{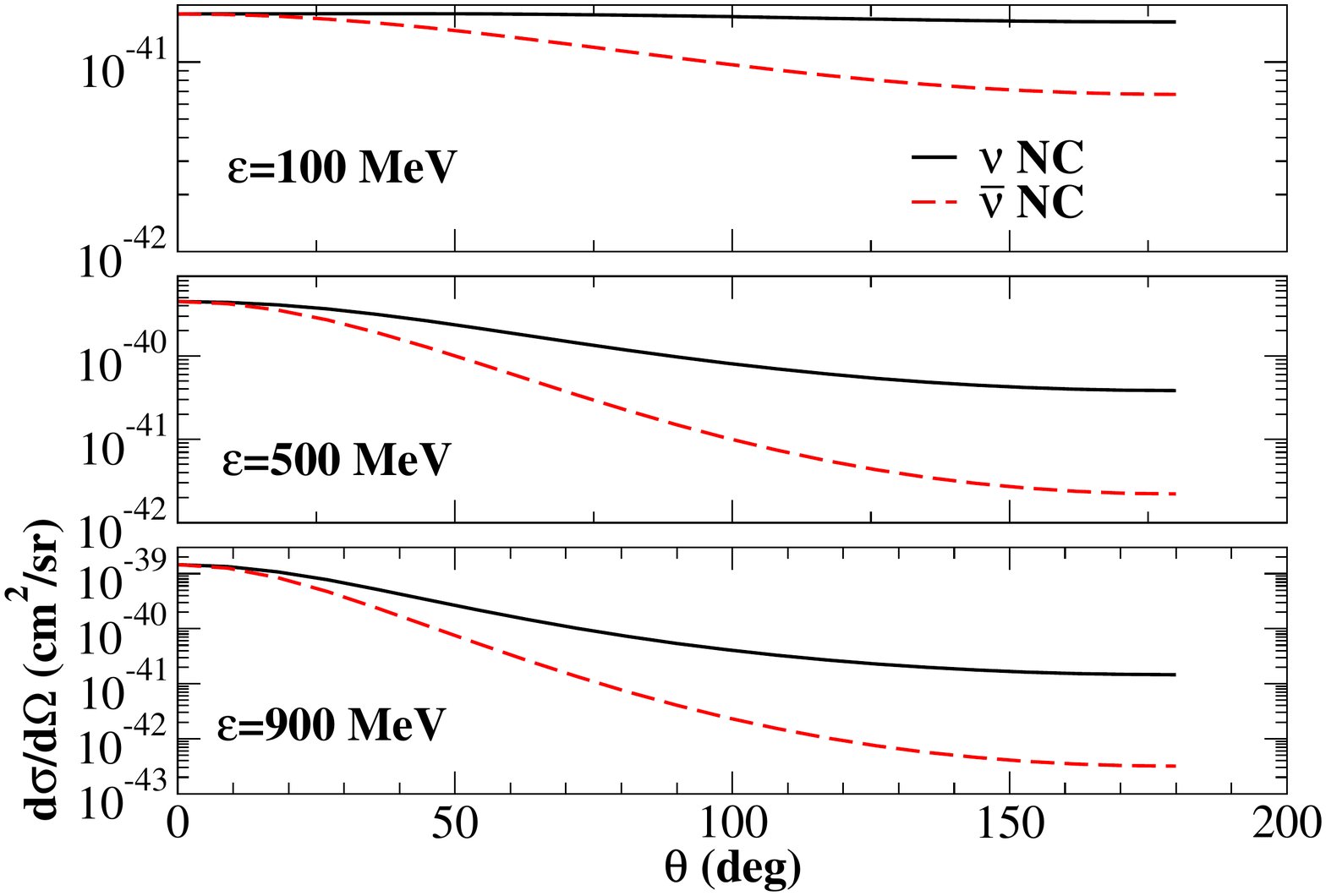}
\includegraphics[width=3.5in]{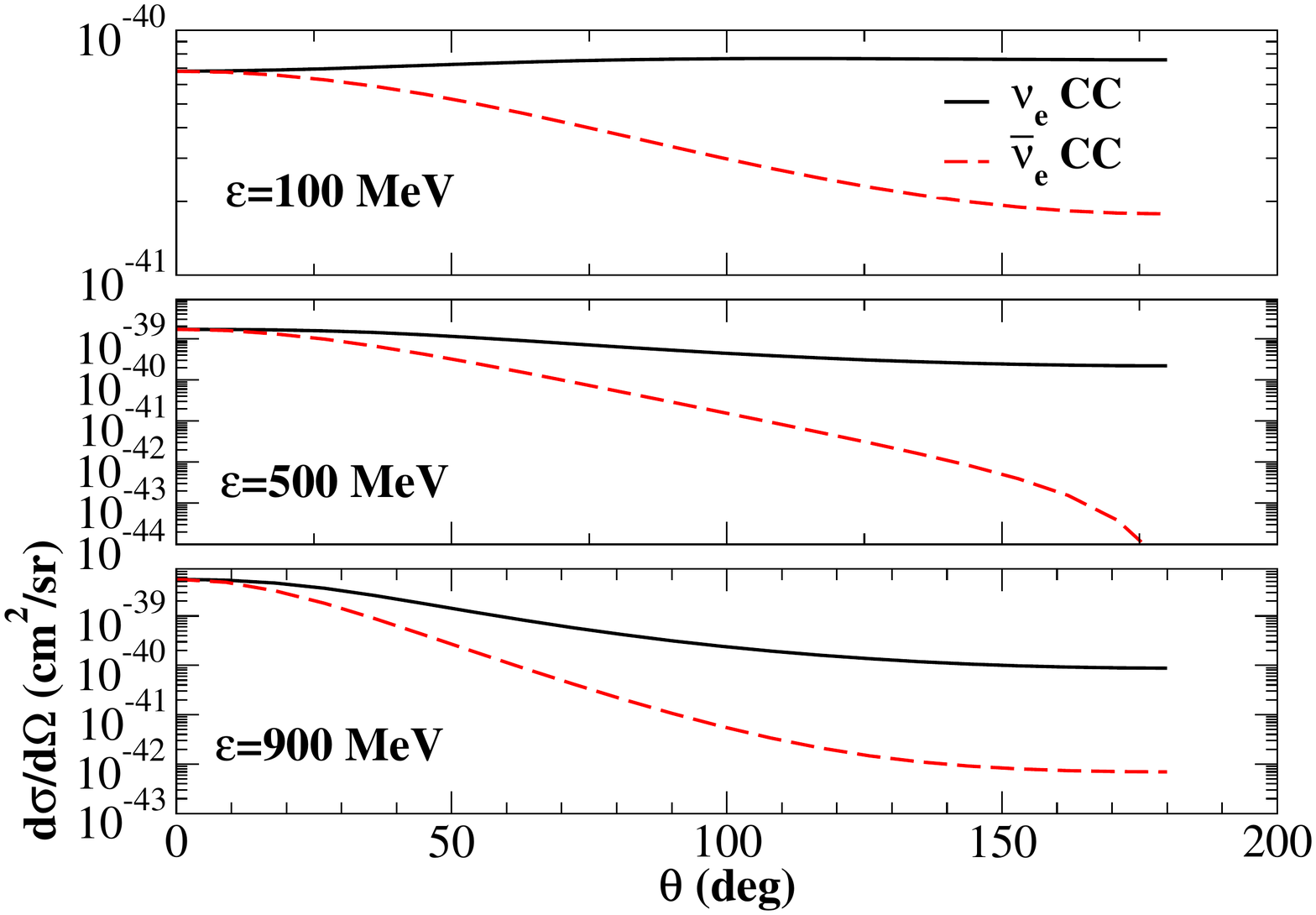}
\caption{(color on line) The ``model'' NC (left panel) and CC (right panel) differential cross sections
for neutrino (solid lines) and antineutrino (dashed lines) energies of 100, 500, and 900 MeV,
as functions of the final lepton scattering angle.}
\label{fig:model1}
\end{figure}

In the naive model, the $\nu$- and $\overline{\nu}$-deuteron NC cross sections are simply
given by the sum of the corresponding proton and neutron (NC) cross sections, while the
$\nu$-deuteron ($\overline{\nu}$-deuteron) CC cross section is identified with the
$n(\nu_e,e^-)p$ [$\,p(\overline{\nu}_e,e^+)n\,$] cross section.  The ``model'' differential cross sections
as function of the final lepton scattering angle (after integrating out the energy-conserving
$\delta$-function) are illustrated in Fig.~\ref{fig:model1} at three incident 
energies ($\epsilon=100, 500, 900$ MeV).  The $\nu$ and $\overline{\nu}$ cross sections
are about the same at forward angles, for which $Q^2$ is small; at backward angles,
as $\epsilon$ and $Q^2$ increase, they both decrease due to the fall off in the form factors. 
However, this fall off is much more pronounced (orders of magnitude) for the $\overline{\nu}$ than
for the $\nu$ cross sections. (At low energy 100 MeV, the form factors do not change
much with angle and the variation with angle in the differential cross section is mild,
still it decreases more in the $\overline{\nu}$ than in $\nu$ channel.) 
These features are reflected in the deuteron cross sections
displayed in Figs.~\ref{fig:nc100}--\ref{fig:pp900} (incidentally, in each panel of
these figures the ``model'' cross sections would be represented by a $\delta$-function
located at $\epsilon^\prime_{\rm qe}$, corresponding to the energy of the quasi-elastic peak).
 
\begin{figure}[bth]
 \centering
\includegraphics[width=6in]{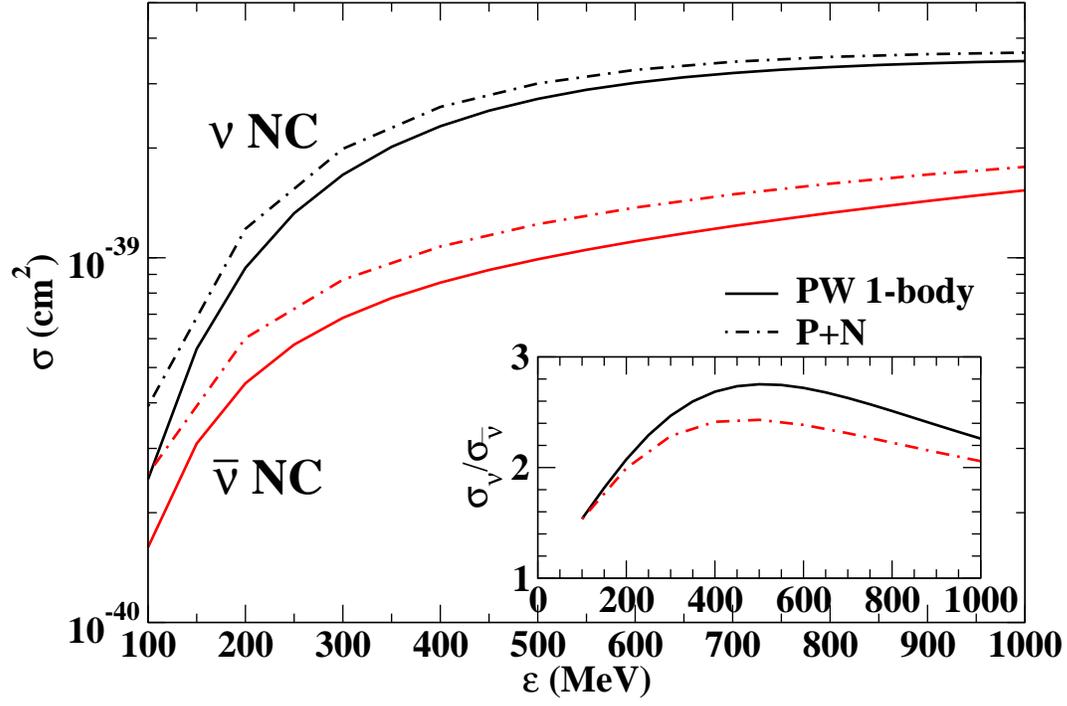}
\caption{(color online) The ``model'' (P+N) NC cross sections for neutrino and antineutrino are compared with plane-wave
one-body (PW 1-body) results, see text for explanation. Inset: ratio of neutrino NC versus antineutrino NC cross section.}
\label{fig:modelnc}
\end{figure}

\begin{figure}[bth]
 \centering
\includegraphics[width=6in]{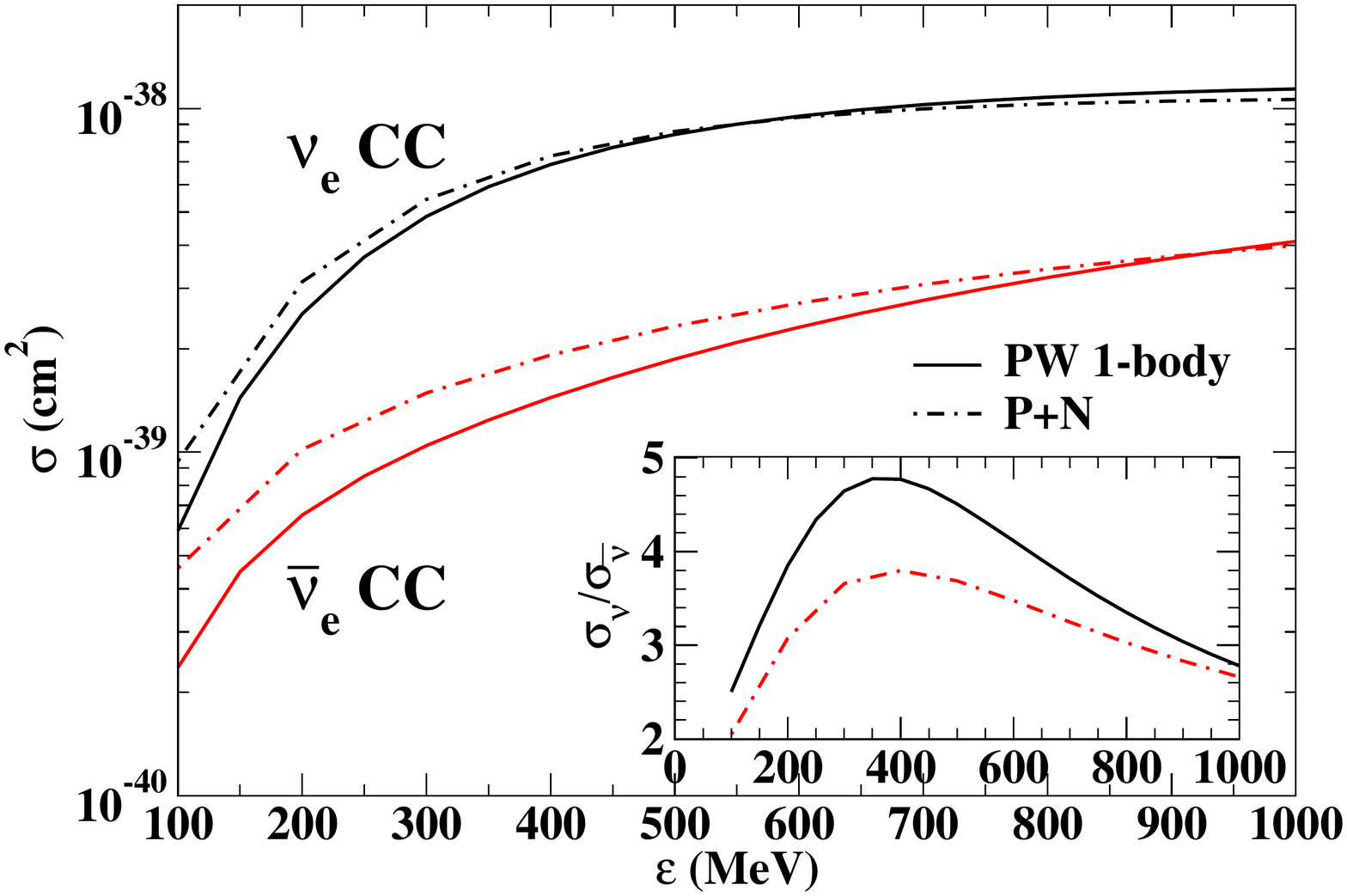}
\caption{(color online) Same as Fig.~\protect\ref{fig:modelnc}, but for CC cross sections.}
\label{fig:modelcc}
\end{figure}

In order to illustrate nuclear correlation effects in the initial deuteron state, we compare the ``model''
$\nu$ and $\overline{\nu}$ NC cross sections with the plane-wave one-body results, shown
in Fig.~\ref{fig:modelnc}, for which we use the physical deuteron state, plane waves for the
two-nucleon continuum states, and one-body currents.  In both $\nu$ and $\overline{\nu}$ NC reactions,
inclusion of nuclear correlations in the initial state reduces the cross sections
from the naive model.  In fact, a similar reduction in cross section (due to nuclear correlations) at about nuclear
density for uniform nuclear matter has been noticed, for example in Refs.~\cite{Horowitz91,Reddy99}.  However,
these correlations increase the ratio of $\nu$ to $\overline{\nu}$ NC cross sections, as shown
in the inset of Fig.~\ref{fig:modelnc}.  Similar effects are also found in the $\nu$ and $\overline{\nu}$
CC reactions at low neutrino energy, as shown in Fig.~\ref{fig:modelcc}.  At higher energies, nuclear
correlations hardly affect these cross sections, and the naive and realistic models are in better
agreement with each other.  The ratio of $\nu$ to $\overline{\nu}$ CC cross sections is also increased
due to nuclear correlations (see inset of Fig.~\ref{fig:modelcc}).  This fact may have interesting implications for long
baseline neutrino experiments aimed at extracting CP violating signals from the detection
of differences in the neutrino and antineutrino channels.


\section{Conclusions and outlook}
\label{sec:concl}

In this work, we have studied inclusive neutrino scattering on the deuteron
up to neutrino energies of 1 GeV, by using a realistic description of two-nucleon
interactions and weak currents.  Two-body terms in the latter increase the
calculated cross sections with one-body currents by less than 10\% over the whole
energy region for both the NC- and CC-induced processes.  Interaction effects in
the two-nucleon continuum final state are found to be negligible for neutrino
energies $\gtrsim 500$ MeV.  This suggests that fairly realistic estimates
for these cross sections in light nuclei (and at relatively high neutrino
energies) may be obtained in calculations based on the plane-wave impulse
approximation.  Even calculations in this limit, however, cannot be presently
carried out, as they require knowledge of nuclear spectral functions
over a wide range of missing momenta and energies---these are not
yet available in light nuclei.  Nuclear correlation effects in the initial deuteron
state are found to be important.  They reduce the $\nu$ and, to a larger extent,
$\overline{\nu}$ cross sections over the whole range of energies studied in this
work, and therefore significantly increase the $\nu$ to $\overline{\nu}$
cross-section ratio for both NC and CC reactions. In the present work the pion-production 
channels are not included. Experimentally they produce distinctive final states and give 
important contributions to total neutrino cross sections above pion-production threshold. 
It would be interesting to include these channels in future.

It should be possible to use quantum Monte Carlo (QMC) methods~\cite{Carlson92}
to study neutrino response functions, and associated sum rules, in light nuclei
within the same (realistic) dynamical framework adopted here.  Indeed, ``exact''
calculations of this type~\cite{Carlson02a} led to a quantitatively 
accurate description of the quasi-elastic electromagnetic response functions measured
in $A=3$ and 4 nuclei.  In particular, they showed that the charge-exchange character
of the nucleon-nucleon interaction leads to shifts of longitudinal
and transverse strength at higher excitation energies, thus providing
a quenching of the response in the quasi-elastic peak region.  This
mechanism, however, is more than offset in the transverse channel by two-body
currents, in particular those associated with pion exchange, and hence
the response is enhanced over the entire quasi-elastic spectrum.
It will be interesting to see the extent to which these considerations
will remain valid in the weak sector probed in neutrino scattering,
and possibly provide an explanation for the observed anomaly in the
$^{12}$C data.

\acknowledgments

It is a pleasure to thank S.\ Nakamura for discussions and clarifications
in reference to his work on the same topic.
The work of R.S. is supported by the U.S. Department of Energy, Office
of Nuclear Science, under contract DE-AC05-06OR23177.
This work was supported in part by a grant from
the DOE under contract DE-AC52-06NA25396.
The calculations were made possible by grants of computing time
from the National Energy Research Supercomputer Center.

\appendix
\section{}
\label{app:app1}

The cross section for CC processes at small incident neutrino
energies in which the lepton mass cannot be neglected reads
\begin{equation}
\left(\frac{ {\rm d}\sigma}{ {\rm d}\epsilon^\prime {\rm d}\Omega}\right)_{\nu/\overline{\nu}}=
 \frac{G^2}{8\,\pi^2}\, \frac{k^\prime}{ \epsilon} \,F(Z,k^\prime)\, \Bigg[ v_{00}\, 
R_{00} +v_{zz} \, R_{zz} -v_{0z}\, R_{0z} +
v_{xx+yy}\, R_{xx+yy} \mp v_{xy} \, R_{xy}
\Bigg] \ ,
\label{eq:xswa}
\end{equation}
where the kinematical factors are given by
\begin{eqnarray}
v_{00}&=& 2\, \epsilon\, \epsilon^\prime\,\left( 1+\frac{k^\prime}{\epsilon^\prime}\, {\rm cos}\, \theta\right)  \ , \\
v_{zz}&=& \frac{\omega^2}{q^2}\left[ m_l^2+ 2\, \epsilon\, \epsilon^\prime\,
\left( 1+\frac{k^\prime}{\epsilon^\prime}\, {\rm cos}\, \theta\right) \right] 
+ \frac{m_l^2}{q^2}\left[ m_l^2+2\,\omega\, (\epsilon+\epsilon^\prime)+q^2\right] \ ,\\
v_{0z}&=& \frac{\omega}{q}\left[ m_l^2+ 2\, \epsilon\, \epsilon^\prime\,
\left( 1+\frac{k^\prime}{\epsilon^\prime}\, {\rm cos}\, \theta\right) \right] +
m_l^2\, \frac{\epsilon+\epsilon^\prime}{q} \ , \\
v_{xx+yy}&=&  Q^2+\frac{Q^2}{2\, q^2}\left[ m_l^2+ 2\, \epsilon\, \epsilon^\prime\,
\left( 1+\frac{k^\prime}{\epsilon^\prime}\, {\rm cos}\, \theta\right) \right] 
-\frac{m_l^2}{q^2} \left[ \frac{m_l^2}{2}+\omega\, \left(\epsilon^\prime+\epsilon\right)\right]\ , \\
v_{xy} &=& Q^2\, \frac{\epsilon+\epsilon^\prime}{q}-m_l^2\, \frac{\omega}{q} \ ,
\end{eqnarray}
$m_l$ is the final lepton mass, and 
the response functions are defined as in Eqs.~(\ref{eq:r1})--(\ref{eq:r5}).
Note that
\begin{equation}
\epsilon+\epsilon^\prime=
\sqrt{2\, m_l^2+({\bf k}+{\bf k}^\prime )^2+Q^2} \ ,
\end{equation}
and the cross section above is easily shown to reduce to Eq.~(\ref{eq:xsw}) in the
limit $m_l=0$ and $Q^2=4\, \epsilon\,\epsilon^\prime
\,{\rm sin}^2\, \theta/2$.

\section{}
\label{app:app2}
In this appendix, the structure functions entering the NC- and CC-induced
processes on the nucleon are expressed in terms of (nucleon) form factors.  In the NC case, they read:
\begin{eqnarray}
A^{\rm NC}&=& 4\, \eta\, \left[ (1+\eta )\, \left(\overline{F}^N_A\right)^2
-(1-\eta) \, \left( \overline{F}_1^N\right)^2 +\eta\, (1-\eta) \left( \overline{F}_2^N\right)^2
+4\, \eta  \, \overline{F}_1^N  \overline{F}_2^N \right] \ ,\\
B^{\rm NC}&=&4\, \eta \, \overline{F}^N_A \left( \overline{F}_1^N + \overline{F}_2^N\right) \ , \\
C^{\rm NC}&=&\frac{1}{4}\left[  \left(\overline{F}^N_A\right)^2+\left( \overline{F}_1^N\right)^2
+\eta\, \left( \overline{F}_2^N\right)^2 \right] \ ,
\end{eqnarray}
and $\eta=Q^2/(4\, m^2)$.  The nucleon form factors $\overline{F}^N_i$ and $\overline{F}^N_A$
for $N=p$ or $n$ are defined as
\begin{eqnarray}
2\, \overline{F}^{\, p/n}_i&=&( 1-4\, \sin^2\theta_W)\, F^{p/n}_i -F^{n/p}_i \ , \\
2 \, \overline{F}^{\, p/n}_A& =& \mp \, G_A \ ,
\end{eqnarray}
where the proton and neutron electromagnetic form factors are,
respectively, $F^p_i =(F_i^S+F_i^V)/2$ and
$F^n_i =(F_i^S-F_i^V)/2$ with $F^{S/V}_i$ defined in Eqs.~(\ref{eq:f1ff})--(\ref{eq:f2ff}),
and the axial form factor $G_A$ (with -- for $p$ and + for $n$) as defined in Eq.~(\ref{eq:gga}).
In the limit in which the final lepton mass and proton-neutron mass difference are both neglected,
the relations for the $A$, $B$ and $C$ structure functions remain valid for the CC case, provided
\begin{equation}
\overline{F}^N_i \longrightarrow F^V_i \ ,\qquad \overline{F}_A^N \longrightarrow G_A \ .
\end{equation}
\end{document}